\magnification1200


\vskip 2cm
\centerline
{\bf   $E_{11}$, Brane Dynamics and Duality Symmetries}
\vskip 1cm
\centerline{ Peter West}
\centerline{Department of Mathematics}
\centerline{King's College, London WC2R 2LS, UK}
\vskip 2cm
\leftline{\sl Abstract}
Following hep-th/0412336 we use the non-linear realisation of the semi-direct product of $E_{11}$ and its vector representation  to construction brane dynamics. The brane moves through a spacetime which arises in the non-linear realisation from the  vector representation and  it contains the usual embedding coordinates as well as the world volume fields. The resulting equations of motion are first order in derivatives and can be thought of as duality relations.  Each  brane carries the full $E_{11}$ symmetry and so the Cremmer-Julia duality symmetries. We  apply this theory  to find the dynamics of  the IIA and IIB strings,  the M2 and M5 branes,  the IIB D3 brane as well as the  one and  two   branes in  seven dimensions.  

\vskip2cm
\noindent

\vskip .5cm

\vfill
\eject
{\bf 1 Introduction}
\medskip
It was conjectured in 2001 that the underlying theory of strings and branes should have an $E_{11}$ symmetry and this should be encoded in a non-linear realisation [1]. The precise conjecture being that the non-linear realisation of 
the semi-direct product of $E_{11}$ with its vector representation, denoted $E_{11}\otimes_s l_1$, is the low energy effective action of strings and branes [2]. Although partial results on the construction of this non-linear realisation were found over the years it was only recently that it was shown that the equations of motion were essentially unique and were those of the maximal supergravity theories if one suitably restricted the fields and coordinates to be those  at the lowest levels [3,4]. Furthermore it has been shown that the non-linear realisation $E_{11}\otimes_s l_1$ is a unified theory in that it contains all the  maximal supergravity theories. The theories in the different dimensions arise from taking  different decompositions of $E_{11}$  and the gauged supergravity theories arise when one considers certain fields to have expectation values. For an review  and the further references the reader is referred to reference  [5]
\par
The vector, or $l_1$,  representation that was used in the above non-linear realisation contains all the known brane charges [2,6,7,8] including those in low dimensions such as two and three. The low level elements of the vector representation begin with the charges of the simple branes which by definition have charges whose spacetime indices are totally antisymmetric,  for example a simple p-brane has a  charge that is of the form $Z^{\underline a_1\ldots \underline a_p}$. However, at higher levels  one finds charges that carry a more complicated index structures,  for example in eleven dimensions the vector representation has, in increasing order of level,  the elements [2]
$$
P_a, Z^{ab}, \ Z^{a_1\ldots a_5}, \ Z^{a_1\ldots a_7,b},\  Z^{a_1\ldots a_8},\ 
\  Z^{ b_1 b_2  b_3,  a_1\ldots  a_8},\ 
\  Z^{( c  d ),  a_1\ldots  a_9},\ 
\  Z^{ c d, a_1\ldots  a_9},\ 
$$
$$
\  Z^{ c, a_1\ldots  a_{10}}\ (2),\ 
Z^{a_1\ldots a_{11}} ,\ 
Z^{ c,  d_1\ldots  d_4, a_1\ldots  a_9},\ 
Z^{ c_1\ldots  c_6, a_1\ldots  a_8},\ 
Z^{ c_1\ldots  c_5, a_1\ldots  a_9},\ 
Z^{ d_1, c_1  c_2  c_3, a_1\ldots  a_{10}},\ (2),
$$
$$
\  Z^{ c_1 \ldots  c_4, a_1\ldots  a_{10}},\ (2),\  Z^{( c_1 c_2, c_3 ),a_1\ldots a_{11}},\ Z^{ c, a_1 a_2},\ (2),\  
Z^{ c_1\ldots  c_{3},a_1\ldots a_{11}},\ (3),\ 
 \ldots  
\eqno(1.1)$$
\par
The blocks of indices contain indices that are totally antisymmetrised while  $()$ indicates that the indices are symmetrised. The elements have  multiplicity one except when there is a bracket after the object which contains a number that gives the multiplicity. All the generator belong to irreducible representations of SL(11), for example $Z^{a_1\ldots a_7,b}$ obeys the constraint $Z^{[ a_1\ldots a_7,b ]}=0$. 
\par
The form charges in the $l_1$ representation, that is, charges that have a single block of totally antisymmetrised indices can be readily computed and are listed in the table below [7,8,9].  
\medskip
\eject
{\centerline{\bf {Table 1. The form generators  in the $l_1$ representation  in D
dimensions}}}
\medskip
$$\halign{\centerline{#} \cr
\vbox{\offinterlineskip
\halign{\strut \vrule \quad \hfil # \hfil\quad &\vrule Ê\quad \hfil #
\hfil\quad &\vrule \hfil # \hfil
&\vrule \hfil # \hfil Ê&\vrule \hfil # \hfil &\vrule \hfil # \hfil &
\vrule \hfil # \hfil &\vrule \hfil # \hfil &\vrule \hfil # \hfil &
\vrule \hfil # \hfil &\vrule#
\cr
\noalign{\hrule}
D&G&$Z$&$Z^{{\underline a}}$&$Z^{{\underline a}_1{\underline a}_2}$&$Z^{{\underline a}_1\ldots {\underline a}_{3}}$&$Z^{{\underline a}_1\ldots {\underline a}_
{4}}$&$Z^{{\underline a}_1\ldots {\underline a}_{5}}$&$Z^{{\underline a}_1\ldots {\underline a}_6}$&$Z^{{\underline a}_1\ldots {\underline a}_7}$&\cr
\noalign{\hrule}
8&$SL(3)\otimes SL(2)$&$\bf (3,2)$&$\bf (\bar 3,1)$&$\bf (1,2)$&$\bf
(3,1)$&$\bf (\bar 3,2)$&$\bf (1,3)$&$\bf (3,2)$&$\bf (6,1)$&\cr
&&&&&&&$\bf (8,1)$&$\bf (6,2)$&$\bf (18,1)$&\cr Ê&&&&&&&$\bf (1,1)$&&$
\bf
(3,1)$&\cr Ê&&&&&&&&&$\bf (6,1)$&\cr
&&&&&&&&&$\bf (3,3)$&\cr
\noalign{\hrule}
7&$SL(5)$&$\bf 10$&$\bf\bar 5$&$\bf 5$&$\bf \overline {10}$&$\bf 24$&$\bf
40$&$\bf 70$&-&\cr Ê&&&&&&$\bf 1$&$\bf 15$&$\bf 50$&-&\cr
&&&&&&&$\bf 10$&$\bf 45$&-&\cr
&&&&&&&&$\bf 5$&-&\cr
\noalign{\hrule}
6&$SO(5,5)$&$\bf \overline {16}$&$\bf 10$&$\bf 16$&$\bf 45$&$\bf \overline
{144}$&$\bf 320$&-&-&\cr &&&&&$\bf 1$&$\bf 16$&$\bf 126$&-&-&\cr
&&&&&&&$\bf 120$&-&-&\cr
\noalign{\hrule}
5&$E_6$&$\bf\overline { 27}$&$\bf 27$&$\bf 78$&$\bf \overline {351}$&$\bf
1728$&-&-&-&\cr Ê&&&&$\bf 1$&$\bf \overline {27}$&$\bf 351$&-&-&-&\cr
&&&&&&$\bf 27$&-&-&-&\cr
\noalign{\hrule}
4&$E_7$&$\bf 56$&$\bf 133$&$\bf 912$&$\bf 8645$&-&-&-&-&\cr
&&&$\bf 1$&$\bf 56$&$\bf 1539$&-&-&-&-&\cr
&&&&&$\bf 133$&-&-&-&-&\cr
&&&&&$\bf 1$&-&-&-&-&\cr
\noalign{\hrule}
3&$E_8$&$\bf 248$&$\bf 3875$&$\bf 147250$&-&-&-&-&-&\cr
&&$\bf1$&$\bf248$&$\bf 30380$&-&-&-&-&-&\cr
&&&$\bf 1$&$\bf 3875$&-&-&-&-&-&\cr
&&&&$\bf 248$&-&-&-&-&-&\cr
&&&&$\bf 1$&-&-&-&-&-&\cr
\noalign{\hrule}
}}\cr}$$
\medskip
 At level zero the $l_1$ representation has the usual spacetime translations $P_{\underline a}$. However,   at level one we find coordinates which are scalars under the SL(D) transformations of our usual spacetime, and so also Lorentz transformations,  but belong to non-trivial representations of $E_{11-D}$. In particular, examining the table we find that they belong to the 
 $$
10,\quad\overline {16}, \quad \overline {27},\quad  56, \quad {\rm  and}\quad 
248\oplus 1 , \quad {\rm of }\quad SL(5),\quad  SO(5,5),\quad 
E_6, \quad  E_7\quad {\rm  and}\  E_8 
\eqno(1.2)$$
 for $  D= 7,6,5,4$ and $3$    dimensions  respectively [8,10].
\par
 Looking at equation (1.1) we see that, except for those at low levels,  the branes charges in the vector representation of $E_{11}$  generically have a more complicated index structure in that they have  more than one block of totally antisymmetrised indices. We will refer to such branes as  exotic, that is, are not simple branes. 
The existence of exotic branes  was first observed in reference [10] which considered the U duality multiplets that contain some of the  well  known simple brane charges. In four and three dimensions, for example, they  found that the  brane charges belonged to rather large multiplets of the U duality symmetry. These authors also postulated a formula for the tensions of the branes that they considered.  Using this  they were able surmise that if the branes could arise by some kind of dimensional reduction  from  some higher dimensional theory, which was unknown at that time,   what would be the index structure of the corresponding brane charges in the higher dimensional theory.  In doing so they came across exotic brane charges. The multiplets found in reference [10] were later shown to be contained in the vector representation of $E_{11}$ [8,9]. 
\par
Thus $E_{11}$ and in particular its $l_1$ representation,  predicts the existence of a very large, in fact an infinite number, of new branes charges and so branes  whose physical 
role has yet to be properly identified [2,7,8,9]. A discussion of how to find what brane   charges are contained in the vector representation, including explicit examples
such as the point particle, string and membrane multiplets in various dimensions was given in reference [7]. Each brane charge in the vector representation corresponds to a weight in this representation and it was shown how one can construct a tension  from the weight and so a tension for the corresponding brane. This formula agrees with the  tensions of the branes where previously known. The dependence on the string coupling could  be read off from one of the components in the corresponding weight in the vector representation [7]. 
\par
In addition to knowing the brane tensions one also has some knowledge of the corresponding solutions. Indeed, given any positive root $\alpha$ in $E_{11}$ one can construct a specific $E_{11}$ group element  [8]
$$
g= e^{-{1\over \alpha^2}(\ln N )\alpha\cdot H}e^{(1-N)E_\alpha}
\eqno(1.3)$$
where  $N$ is a function whose form is not specified. From  the group element one can read off the values of the fields and so a putative solution.  Using  low level roots in eleven and ten dimensions one finds all the half BPS branes in these dimensions. In fact one can apply this formula to find generic solutions for any positive $E_{11}$ root and in any dimension. Thus one finds a large number of new putative solutions whose charges will not generically be those of simple branes [8]. This construction was generalised to a formula for a group element that depends on two positive roots in eleven [11] and ten dimensions  [12]. These solutions for low level roots reproduced all the quarter BPS branes in these dimensions. A further generalisation to include three and more  numbers of $E_{11}$ roots was given in refence [11]. 
\par
When a rigid symmetry group $G$ of a quantum field theory is spontaneously broken to a subgroup $H$, Goldstone theorem tells us the number of resulting massless particles. Furthermore,  the low energy action that describes these particles is almost always the non-linear realisation of the group  $G$ with local subgroup $H$. This was the approach used to compute the dynamics of pions in the early days of particle physics. In this application spacetime was introduced by hand as a variable that the fields in the non-linear realisation depended on.  Much  of the past literature on $E_{11}$ has been devoted to constructing the $E_{11}\otimes l_1$ non-linear realisation to construct a field theory, see for example in references [3] and [4],  and it is in this theory that the maximal supergravity theories are contained.  In contrast to the non-linear realisations used to describe  pion dynamics,  the 
$E_{11}\otimes l_1$ non-linear realisation automatically encodes a spacetime as its coordinates arise as the coefficients of the generators of the vector representation as they occur in the group element used to construct the non-linear realisation. 
\par
However, one can also use non-linear realisations to derive brane dynamics. 
An incomplete list of some of these papers is given in reference [13]. The bosonic p-brane in $D$ dimensions can be thought of as the non-linear realisation of $SO(1,D-1) \otimes T^D$,  where $T^D$ are generators in the vector representation, and  the local subgroup is 
$SO(1,D-p-2)\otimes SO(1,p)$ [14]. To construct the brane dynamics in the presence of background gravity fields one only has to  consider the non-linear realisation of the group $GL(1,D-1)\otimes_s T^D$ [15].
\par
Reference [9] outlined how to construct  brane dynamics  using the non-linear realisation of $E_{11}\otimes_s l_1$ and in particular gave a partial construction of the dynamics of the M2 and M5 branes. 
In this approach the coordinates that arise from the generators of the vector representations describe the embedding of the brane in the background spacetime, but  they are taken to be functions of the parameters that parameterize  the brane world volume. Since the vector representations contains all brane charges the brane coordinates arise from the brane charges in a natural way in this construction. The coupling of the brane to the supergravity fields is also automatic as these fields occur in the non-linear realisation as the coefficients of the Borel sub algebra generators of in the $E_{11}$ group element. 
\par
The $E_{11}\otimes_s l_1$ non-linear realisation was used to construct the dynamics of the IIA string [16] by taking the decomposition of $E_{11}$ that leads to the IIA theory in ten dimensions. The result was a SO(10,10) invariant formulation of the string whose coordinates belonged to the vector representation of SO(10,10). The final result agreed with the previously found  result of references [17] and [18].  Reference [19] realised that to extend this work to consider other types of branes and encode higher duality symmetries required a new way of thinking and more recently  reference [20] also raised the question of how one could incorporate the well known Cremmer-Julia symmetries into the brane dynamics.   
\par
In this paper we will extend the discussion of reference [16]  and further  develop the $E_{11}\otimes_s l_1$ non-linear realisation as it can be applied to branes in section two. In section three we discuss the   dynamics of  IIA string and in section four the dynamics of the  the M2 brane and the  M5. In section five we construct the IIB string and the D3 brane. In section six we consider  the one brane and two brane dynamics in seven dimensions.  Finally, in section seven  discuss some general features of brane dynamics that emerge from the non-linear realisation.  The calculations in  the IIB theory and in seven dimensions use  the Cartan involution invariant subalgebra of $E_{11}\otimes_s l_1$ in the decomposition appropriate to these  theories; these results will  be published elsewhere [33]. 
\par
The main aim of this paper is to further   developing  the $E_{11}\otimes_s l_1$ non-linear realisation in the hope that we may use it to  compute brane dynamics for all the branes in E theory. As we show it does lead to dynamics of some of the well known branes and gives the dynamics of the new branes in seven dimensions.  This approach automatically encodes the Cremmer-Julia symmetries as they appear at level zero in $E_{11}$. 
\medskip
{\bf 2. General Formalism}
\medskip
We are interested in the semi-direct product of $E_{11}$ with its vector representation $l_1$, which we denoted by $E_{11}\otimes_s l_1$. The commutators of this algebra  can be written in the form 
$$
[R^{\bar \alpha} , R^{\bar \beta} ]= f^{\bar \alpha \bar \beta}{}_{\bar \gamma} R^{\bar \gamma}, \quad
[R^{\bar \alpha} , l_A]= -(D^{\bar \alpha} )_A{}^B l_B
\eqno(2.1)$$
where $R^{\bar \alpha}$ are the generators of $E_{11}$ and $l_A$ are the generators belonging to the vector ($l_1$) representation. We assume that the $l_1$ generators commute. The matrices $(D^{\bar \alpha} )_A{}^B$ are the representation matrices of the $E_{11}$ algebra in the $l_1$ representation. In previous papers we have used $\underline \alpha$ for the indices of the $E_{11}$ generators but in this paper we will use $\bar \alpha$. 
\par
An important part will be played by the Cartan involution invariant subalgebra of $E_{11}$ which we denote by $I_c(E_{11})$. The Cartan involution $I_c$  takes positive root generators to negative root generators and its action can be taken to be 
$$
I_c(R^{\bar \alpha}) = - R^{-\bar \alpha} 
\eqno(2.2)$$ 
for any $E_{11}$ root $\bar\alpha$. The Cartan involution subalgebra is generated by 
$S^{\bar \alpha} \equiv R^{\bar \alpha} - R^{-\bar \alpha} $. 
\par
A non-linear realisation is specified by the choice of an  algebra together with a choice of  subalgebra, referred to as the local subalgebra. The non-linear realisation which leads to brane dynamics is a non-linear  realisation of $E_{11}\otimes_sl_1$  with a local subalgebra  ${\cal H}$ that is a subgroup of $I_c(E_{11})$. The different choices of subalgebra ${\cal H}$  lead to the different branes. 
\par
\par
The non-linear realisation is  constructed from  a group element  $g\in E_{11}\otimes_sl_1$ and we have to construct an action, or set of equations of motion,  that is  invariant under the transformations 
$$
g\to g_0 g, \ \ \ g_0\in E_{11}\otimes _s l_1,\ \ {\rm as \  well \  as} \ \ \ g\to gh, \ \ \ h\in
{\cal H}
\eqno(2.3)$$
The group element $g_0\in E_{11}\otimes_s l_1$ is a rigid transformation, that is, it is  a constant,  while the group element $h$ belongs to the local subalgebra ${\cal H}$ and it   is a local transformation whose precise meaning will be discussed just below. 
\par
We can write the group element $g$ of the  non-linear realisation  in the form 
$$
g=g_l g_h g_E 
\eqno(2.4)$$
In this equation  $g_E$ is in the Borel subgroup   of $E_{11}$,  the group element   $g_l$ is formed from the generators of the $l_1$ representation
while the group element $g_h$ belongs to $I_c(E_{11})$. 
 We can write the individual group elements in the form 
$$
g_l= e^{z^A l_A}, \quad g_E=e^{A_{\bar \alpha} R^{\bar \alpha}}, 
\quad g_h= e^{\varphi \cdot S}
\eqno(2.5)$$
In  this equation the parameters $z^A$ are the coordinates of the background space-time  and they depend on the coordinates of the brane world volume  $\xi^{\underline \alpha}$. The  
$A_{\underline\alpha}$ are   the $E_{11}$ background fields, which include those of the maximal supergravity theories,    and they depend on the coordinates of the background spacetime
$z^A$. The fields $\varphi$ also depend on  $\xi^{\underline \alpha}$ and by a local transformation we mean one that depends on $\xi^{\underline \alpha}$. Clearly, we may use the local subalgebra ${\cal H}$ to set some  of the $\varphi$  fields  to zero. The brane world volume coordinates include,  at lowest level,  the usual brane coordinates $\xi^{\alpha}$ but they may also contain the higher level coordinates, for example in eleven dimensions the next possible coordinates would be $\xi_{\alpha_1\alpha_2}$. The presence of the higher level coordinates corresponds to the possibility that the brane moves not only in the usual spacetime but also part of the background spacetime $z^A$. For branes with no world volume fields we will not need the higher level brane world volume coordinates, but they seem to be required when world volume fields are present. 
\par
It is apparent from this way of constructing brane dynamics that a given brane  will be invariant under all the $E_{11}\otimes_s l_1 $ symmetries, but which of these symmetries are  linearly realised and which are spontaneously broken, and so non-linearly realised,  depends on the choice of local subalgebra ${\cal H}$ which in turn  depends on the brane we are studying. We note that  every brane will automatically be invariant under all the supergravity $E_{11-d}$ duality symmetries in $d$ dimensions as these occur in $E_{11}$ at the lowest level. 
\par
As we have mentioned the dynamics is just that invariant under the symmetries of equation (2.3) and the best method to find these equations is to consider the Cartan forms 
$$
{\cal V}= g^{-1} d g= {\cal V}_E+{\cal V}_l +{\cal V}_h, 
\eqno(2.6)$$
where 
$$
{\cal V}_E= g_E^{-1}dg_E , \ {\rm and }\ 
{\cal V}_l= g_E^{-1} g_h^{-1} (g_l^{-1}dg_l) g_h g_E  , \quad {\cal V}_h= g_E^{-1} (g_h^{-1}d g_h ) g_E
\eqno(2.7)$$
Clearly ${\cal V}_E$ belongs to the $E_{11}$ algebra and are just the Cartan forms of $E_{11}$; we can write them   as 
$$
{\cal V}_E \equiv (dz^\Pi G_{\Pi, \underline \alpha} R^{\underline \alpha}) 
\eqno(2.8)$$
where the $G_{\Pi, \underline \alpha}$ just depend on the $E_{11}$ background fields $A_{\underline \alpha}$.  We can  write  
$$
{\cal V}_l\equiv d\xi^{\underline \alpha} \nabla^B _{\underline \alpha} z^A l_A = g_E^{-1}g_h^{-1} (dz^Al_A ) g_h g_E = g_E^{-1} (d\xi^{\underline \alpha} \nabla _{\underline \alpha} z^A  l_A )g_E\equiv    d\xi^{\underline \alpha} \nabla _{\underline \alpha} z^\Pi  E_\Pi{}^Al_A
\eqno(2.9)$$
where $E_\Pi{}^A $ is defined by  $g_E^{-1} dz\cdot l g_E\equiv 
dz^\Pi E_\Pi{}^A l_A  $. This last object just depends on the $E_{11}$ background fields and it is the vielbein in background spacetime with coordinates $z^A$ while $ \nabla _\alpha z^A$ depends only on the coordinates $z^A$ and the fields $\varphi$. 
\par
It is instructive to recall how the above non-linear realisations differs from that used to derive the low energy effective field theory describing  the behaviour of strings and branes.  In this case the local subgroup is $I_c(E_{11})$ and so we may choose the group element $g_h$ to be the identity element and so there are no $\varphi$ fields.  
There is no brane and so   no brane parameters $\xi^{\underline \alpha}$. However, we do have the coordinates $z^A$ and the fields  $A_{\underline \alpha}$ which depend on  these coordinates. As a result, the Cartan forms associated with the vector representation just contain the vielbein and so are  functions of the $E_{11}$ fields. The dynamics is essentially encoded in equations which are functions of the $E_{11}$ Cartan forms 
${\cal V}_E$ and the vielbein is used to convert world to tangent indices on these objects. We recall that the construction leads to a field theories which when suitably truncated to low levels are the maximal supergravity theories. 
\par
We note that in the non-linear realisation used to construct branes, the vielbein defined below equation (2.9) and  the Cartan forms of equation (2.8)  just contain the   $E_{11}$ fields and they are the same as one finds in the non-linear realisation  discussed  in the paragraph just above. As a result computing the non-linear realisation for the brane dynamics, set out above,  will also lead to the low energy effective action for strings and branes whose truncation contains the maximal supergravity theories. The expressions for ${\cal V}_E$ and the vielbein, at low levels, can be found, for example,  in references [3] and [27]. The fields $\varphi$ only  occur in the Cartan forms  $\nabla^B _{\underline \alpha} z^A$, or equivalently $\nabla _{\underline \alpha} z^A$,  associated with the vector representation and 
the dynamics of the branes will consist of invariant equations among these later objects. It is these later equations that we will focus on deriving in this paper. 
\par
In this paper we will   compute  the brane dynamics in the absence of background fields and so we will take $g_E$ to be the identity matrix. In this case the non-linear realisation we are constructing is for the algebra $I_c(E_{11})\otimes_s l_1 $ with local subgroup ${\cal H}$ and so the group element has the form $g=g_lg_h$ and the Cartan forms are given by 
$$
{\cal V}= g^{-1} d g= {\cal V}_l +{\cal V}_h, 
\eqno(2.10)$$
where 
$$
{\cal V}_l= g_h^{-1}(g_l^{-1}dg_l)  g_h = g_h^{-1}( d z^A l_A)  g_h \equiv \nabla_\alpha z^A l_A, \quad {\cal V}_h=  (g_h^{-1}d g_h ) 
\eqno(2.11)$$
Under the rigid transformation $g_0\in I_c(E_{11})$ 
$$
g_l \to g_0 g_l g_0^{-1} ,\quad {\rm or \ \ equivalently } \quad dz^A l_A \to g_0 dz^A l_A g_0^{-1} 
, \quad g_h\to g_0 g_h
\eqno(2.12)$$
while under the local transformation $h\in {\cal H}$ we have 
$$
g_l\to g_l , \quad g_h\to g_h h 
\eqno(2.13)$$
The Cartan forms are inert under the rigid $g_0$ transformations. but under the local $h\in {\cal H}$ transformations they transform as   
$$ 
{\cal V}\to  h^{-1}{\cal V} h + h^{-1} d h
\eqno(2.14)$$
and in particular that 
$$
\nabla z^A l_A \to h^{-1} ( \nabla z^A l_A ) h ,  \quad 
{\cal V}_h\to  h^{-1}{\cal V}_h h + h^{-1} d h
\eqno(2.15)$$
where $d\xi ^{\underline \alpha } \nabla_{\underline \alpha }$. 
Using this equation it is straightforward to explicitly compute these transformations from the $E_{11}\otimes _s l_1$ algebra. The dynamics is a set of equations that are invariant under these transformations and as the $\nabla_{\underline \alpha} z^A $ transform covariantly we are looking for equations constructed from these quantities that transform into each other. We will also demand that the equations are invariant under arbitrary reparameterisations of the brane world volume, that is , diffeomorphism in the parameters $\xi^{\underline \alpha}$. 
\par
The brane dynamics  in the presence of the background fields can be readily found from the resulting equations. In particular,  by using equation (2.9), we can  reinstate their presence by introducing the  vielbein in the way that this equation dictates, that is, make the replacement 
$$
\nabla _{\underline \alpha} z^A \ \to \ \nabla^B _{\underline \alpha }z^A 
\eqno(2.16)$$ 
The veilbein that occurs in the non-linear realisation has been computed in several dimensions in reference [27]. We will not in this paper consider the construction of the Wess-Zumino term in the brane dynamics. 
\medskip
{\bf 3. The IIA  string}
\medskip
The IIA theory arises from the 
 non-linear realisation of $E_{11}\otimes_s l_1$ when we decompose the  $E_{11}$ algebra in  terms of the SO(10,10) algebra that results from deleting the node ten in the $E_{11}$ Dynkin diagram. The 
SO(10,10) group  is  the T duality group of string theory. We recall that, at level zero,  the non-linear realisation of $E_{11}\otimes_s l_1$ leads [21]  precisely to Siegel theory [22,23] which contains the massless fields of the NS-NS sector of string theory. The extension to include the massless fields of the R-R sector of the IIA string theory was found by computing the level one part of the $E_{11}\otimes_s l_1$ non-linear realisation [24]. 
\par
The method of constructing dynamics in  a field theory from a non-linear realisation, usually uses the Cartan forms as they are inert under the rigid symmetries of the non-linear realisation. Indeed the construction of the $E_{11}$ low energy effective action of strings and branes has been  found using this method. 
However, there is another method that works instead with the object $M\equiv g I_c(g^{-1})$ which is inert under the local transformations. Generally this later method is less useful for deriving field theory results and indeed difficult to implement in the $E_{11}$ context. Nonetheless  the dynamics of the IIA string has already been constructed from the non-linear realisation using the quantity  $M$ [16].  The results agreed with that of reference [17]. 
\par
In this section we will use the more conventional method to construct the dynamics of the IIA string using the method of Cartan forms. We will find that this procedure is quite different to the $M$ method and has a number of  usual and unexpected features. As such it is useful to introduce this method in a simple setting so as to clearly illustrate the unusual features before using the method  to construct the dynamics of  more complicated  branes dynamics given  later in this paper. 
\par
 At level zero the  $E_{11}$ generators are $K^a{}_b$, $R^{ab}$ and $\tilde R_{ab}$, while in the $l_1$ representation, we have the generators $P_a$
 and $Q^a$ [19]. The commutators of the semi-direct product algebra $E_{11}\otimes_s l_1$ at level zero are given by [19]
$$
[K^{\underline a}{}_{\underline b},K^{\underline c}{}_{\underline d}]=\delta _{\underline b}^{\underline c} K^{\underline a}{}_{\underline d} - \delta _d^{\underline a} K^{\underline c}{}_{\underline b}, \ \
[K^{\underline a}{}_{\underline b}, R^{{\underline c} {\underline d}}]=\delta_{\underline b}^{\underline c} R^{{\underline a}{\underline d}}-\delta_{\underline b}^{\underline d} R^{{\underline a}{\underline c}},
[K^{\underline a}{}_{\underline b}, \tilde R_{{\underline c} {\underline d}}]=-\delta_{\underline c}^{\underline a} \tilde R_{{\underline b}{\underline d}}+\delta_{\underline d}^{\underline a} \tilde
R_{{\underline b}{\underline c}},
$$
$$
[R^{a{\underline b}}, \tilde R_{{\underline c} {\underline d}}]=\delta_{[{\underline c}}^{[a}
K^{{\underline b}]}{}_{{\underline d}]},\ \ \ [R^{a{\underline b}}, R^{{\underline c} {\underline d}}]=0=[\tilde R_{a{\underline b}}, \tilde R_{{\underline c} {\underline d}}]
\eqno(3.1)$$
$$
[K^{\underline c}{}_{\underline b}, P_{\underline a}]=-\delta_{\underline a}^{\underline c} P_{\underline b},\ \ [ R^{{\underline a}{\underline b}}, P_{\underline c}]=-{1\over 2}
(\delta^{\underline a}_{\underline c} Q^{\underline b}-\delta^{\underline b}_{\underline c} Q^{\underline a}),\ \ [\tilde R_{{\underline a}{\underline b}}, P_{\underline c}]=0,
\eqno(3.2)$$
$$
[K^{\underline a}{}_{\underline b}, Q^{\underline c}]=\delta_{\underline b}^{\underline c} Q^{\underline a},\ \ [ \tilde R_{{\underline a}{\underline b}}, Q^{\underline c}]={1\over
2} (\delta_{\underline a}^{\underline c} P_{\underline b}-\delta_{\underline b}^{\underline c} P_{\underline a}),\ \ [R^{{\underline a}{\underline b}},Q^{\underline c}]=0,
\eqno(3.3)$$
where $\underline a , \underline b , \ldots = 0,1,\ldots ,D-1$.
\par
The Cartan involution acts on the generators in the following way
$$ 
I_c(K^{\underline a}{}_{\underline b}) = -K^{\underline b}{}_{\underline a}, \ \ I_c(R^{{\underline a}{\underline b}}) = - \tilde{R}_{{\underline a}{\underline b}} 
\eqno(3.4)$$
and as a result the  Cartan Involution invariant subalgebra, $I_c(SO(10,10))$ is generated by

$$ 
J_{\underline{a}}{}^{\underline{b}} \equiv K^{\underline{a}}{}_{\underline{b}} - K^{\underline{b}}{}_{\underline{a}}, \ \ S_{\underline{a}\underline{b}} \equiv 2(R^{\underline{a}\underline{b}}
 - \tilde{R}_{\underline{a}\underline{b}})\eqno(3.5) $$
The $I_c(SO(10,10))$  algebra is given by 
$$
[J_{\underline{a} \underline{b}}, J_{\underline{c} \underline{d}}]=  
\eta_{\underline{b} \underline{c}}J_{\underline{a} \underline{d}}
-\eta_{\underline{a} \underline{c}}J_{\underline{b} \underline{d}}
-\eta_{\underline{b} \underline{d}}J_{\underline{a} \underline{c}}
+\eta_{\underline{b} \underline{c}}J_{\underline{a} \underline{d}}
$$
$$
[S_{\underline{a} \underline{b}}, S_{\underline{c} \underline{d}}]=  
\eta_{\underline{b} \underline{c}}J_{\underline{a} \underline{d}}
-\eta_{\underline{a} \underline{c}}J_{\underline{b} \underline{d}}
-\eta_{\underline{b} \underline{d}}J_{\underline{a} \underline{c}}
+\eta_{\underline{b} \underline{c}}J_{\underline{a} \underline{d}}
$$
$$
[J_{\underline{a} \underline{b}}, S_{\underline{c} \underline{d}}]=  
\eta_{\underline{b} \underline{c}}S_{\underline{a} \underline{d}}
-\eta_{\underline{a} \underline{c}}S_{\underline{b} \underline{d}}
-\eta_{\underline{b} \underline{d}}S_{\underline{a} \underline{c}}
+\eta_{\underline{b} \underline{c}}S_{\underline{a} \underline{d}}\eqno(3.6)$$
By adopting suitable generators one sees that it is none other than the algebra $SO(10)\otimes SO(10))$. Their 
commutators with the $l_1$ representation are  given by 

$$ [J_{ab}, P_c] = -2\eta_{c[a}P_{b]}, \ \ [S_{ab}, P_c] = -2\delta_c^{[a}Q^{b]} $$
$$ [J_{ab}, Q^c] = -2\delta^c_{[a}Q_{b]} , \ \ [S_{ab}, Q^c] = - 2\delta^c_{[a}P_{b]}  
\eqno(3.7)$$
\par
We will first consider the brane in the absence of background fields and so we will take the non-linear realisation of $SO(10)\otimes SO(10)\otimes _s l_1$.  
The IIA string lives in ten dimensions but our considerations can  easily 
be generalised to $D$ dimensions and by  consider the non-realisation of $SO(D,D)\otimes _s l_1$ and trivial increase in the range of our indices. 
\par
 We choose the   local subgroup of our   non-linear realisation to be the subgroup whose algebra  is given by 
$$ 
{\cal H} = \{ J_{ab}, S_{ab}, J_{a'b'}, S_{a'b'}, L_{ab'} , \ldots \} 
\eqno(3.8)$$
where $a,b,\ldots =0,1$ and $a^\prime ,b^\prime ,\ldots =2,\ldots ,9$ and 
$L_{ab'} \equiv J_{a b'}-\epsilon_{a}{}^{c} S_{cb^\prime }$. We note that the local subalgebra ${\cal H}$ is not $SO(1,1)\otimes SO(8)\otimes SO(1,1)\otimes SO(8)$ as one might naively expect. The $M$ method,  used in reference [16], is rather insensitive to the precise choice of local subalgebra used in the non-linear realisation and it would not be sensitive to the different choice of local subalgebra. 
\par
Using the commutation relations of equation (3.6)  we find that the commutators of the local subalgebra  ${\cal H}$ to be given by certain of the commutators of equation (3.6) as well as 
$$ 
[L_{ab'}, L_{cd'}] = 0 , \ \ [S, L_{cd'}] = - L_{cd'} 
\eqno(3.9)$$
$$ [ S_{a'b'}, L_{cd'}] = -\eta_{b'd'}\varepsilon_c{}^eL_{ea'} + \eta_{a'd'}\varepsilon_a{}^eL_{eb'} \ \ 
[ J_{a'b'}, L_{cd'}] = +\eta_{b'd'}\varepsilon_c{}^eL_{ea'} -\eta_{a'd'}\varepsilon_a{}^eL_{eb'} 
\eqno(3.10)$$
where $S = \varepsilon_{ab}S$. 
\par
The commutators of the generators of ${\cal H}$ with the vector  representation are given by equation (3.7) as well as   
$$
 [L_{ab'}, P_c] = - \eta_{ac} P_{b^\prime} + \epsilon _{ac}Q_{b^\prime} ,\ 
[L_{ab'}, Q_c] = - \eta_{ac} Q_{b^\prime} + \epsilon _{ac}P_{b^\prime} ,\ 
[S,P_a]=\epsilon _{ab} Q^c , \ , \ [S,P_{a^\prime} ]=0$$
$$
[L_{ab'}, P_{c^\prime }] =  \eta_{b'c'} (P_a - \epsilon_{a}{}^c Q_c),\ 
[L_{ab'}, Q_{c^\prime }] =  \eta_{b'c'} (Q_a - \epsilon_{a}{}^c P_c) . \ 
[S,Q_a]=\epsilon _{ab} P^c , \ [S,Q_{a^\prime} ]=0
\eqno(3.11)$$
\par
It will prove useful to divide the generators of the $l_1$ representation into two sets which are given by 
$$
\{N_c^- \equiv P_c - \varepsilon_{cd}Q_d \}
\eqno(3.12) $$
and  
$$
\{ N_c^+ \equiv  P_c + \varepsilon_{cd}Q^d, P_{a^\prime}, Q_{a^\prime} \}
\eqno(3.13)$$
The commutators of these newly defined generators are  those of equation (3.7) as well as 
$$
 [ L_{ab'}, N^-_c] = 0 ,\  [L_{ab'}, N^+_c] = -2(\eta_{ac}P_{b'} - \varepsilon_{ac}Q_{b'}), \ \ [S, N^\pm_c] = \mp 2 N_c^ \pm 
$$
$$
[ L_{ab'}, P_{c^\prime} ]= \eta_{b^\prime c^\prime} N_a^- , 
\ [ L_{ab'}, Q_{c^\prime} ] = - \eta_{b^\prime c^\prime} \epsilon _{a}{}^{d} N_d^- ,\ [S , P_{b^\prime} ]= 0 , \ [S , Q_{b^\prime} ]= 0
\eqno(3.14)$$
We observe that the generators of the vector representation contain an irreducible representation when decomposed into representations of ${\cal H}$, namely $N_c^- $. We observe that the compliment of this representation does not transform into itself. 
\par
As explained in section two, the non-linear realisation is constructed from the group element $g=g_E g_h$ which we can now write in the form 
$$
 g = e^{x^{\underline{a}}(\xi)P_{\underline{a}} + y_{\underline{a}}(\xi)Q^{\underline{a}}}e^{-\varphi^{ab^\prime} J_{ab^\prime}}
 \eqno(3.15) $$
where,  using the local symmetry of equation (2.13), we have chosen to $g_h$ to be of the above form. We  note that the only generators of $I_c(SO(10,10))$ which are  not in the local subgroup ${\cal H}$ can be taken to be $ J_{ab^\prime}$. Examining the group element we  find that the string moves in the background spacetime with coordinates 
$x^a$ and $y_a$ but we take the world volume coordinates to be just the $\xi^\alpha$. 
\par
The Cartan form  ${\cal V} \equiv g^{-1} d g$ can be written in the form 
$$ {\cal V} = d\xi^\alpha (\nabla_\alpha x^aP_a + \nabla_\alpha y^aQ_a + \nabla_\alpha x^{a'}P_{a'} + \nabla_\alpha y_{a'}Q^{a'} )
\eqno(3.16)$$
Writing this in terms of the generators of equations (3.12) and (3.13) we find it becomes 
 $$
 {\cal V} = d\xi^\alpha ({\cal E}_\alpha {}_-^a  N_a^- +{\cal E}_\alpha {}_+^a N_a^+ + \nabla_\alpha x^{a'}P_{a'} + \nabla_\alpha y_{a'}Q^{a'} )
\eqno(3.17)$$
where 
$$
{\cal E}_\alpha {}_\pm^a \equiv  {1\over2}(\nabla_\alpha x^a\mp\varepsilon^{ab}\nabla_\alpha y_b)
\eqno(3.18)$$
\par
Using  equation (2.15) we finds that under the transforms $h=1+ \Lambda^{ab^\prime} L_{a b^\prime}+\Lambda S$ the variations of the Cartan forms  is given by 
$$
\delta (\nabla_\alpha x^a)
=-\Lambda^{ab^\prime}\nabla_\alpha x_{b^\prime} -\nabla_\alpha  y_{b^\prime}
\epsilon^{ac} \Lambda_{c}{} ^{b^\prime} -\Lambda \nabla_\alpha y_b \epsilon^{ba},\ 
$$
$$
\delta (\nabla_\alpha y^a)
=-\Lambda^{ab^\prime}\nabla_\alpha y_{b^\prime} -\nabla_\alpha  x_{b^\prime}
\epsilon^{ac} \Lambda_{c}{} ^{b^\prime} -\Lambda \nabla_\alpha x_b \epsilon^{ba}, \ 
$$
$$
\delta (\nabla_\alpha x^{b^\prime})
=\Lambda^{ab^\prime}(\nabla_\alpha x_{a} -\epsilon^{ac} \nabla_\alpha  y_{c}) ,\ 
\delta (\nabla_\alpha y^{b^\prime})
=\Lambda^{ab^\prime}(\nabla_\alpha y_{a} -\epsilon^{ac} \nabla_\alpha  x_{c})
\eqno(3.19)$$
When expressed in terms of the  the variables in equation (3.17) the same result is given by  
$$
\delta {\cal E}_\alpha {}_+^a=0 , \ 
\delta {\cal E}_\alpha {}_-^a= -\Lambda ^{ab^\prime} \nabla_\alpha x_{b^\prime} - 
\Lambda ^{db^\prime} \epsilon_{ad} \nabla_\alpha y_{ b^\prime} , \ 
$$
$$
\delta (\nabla_\alpha x^{b^\prime}) = 2\Lambda^{ab^\prime} {\cal E}_\alpha {}_+ ^a ,\ 
\delta (\nabla_\alpha y^{b^\prime}) = -2\epsilon _{ac} \Lambda^{ab^\prime} {\cal E}_\alpha {}_+ ^c  
\eqno(3.20)$$
\par
Examining these variations it is readily apparent that a set of equations that is invariant under the transformations of ${\cal H}$, and so all the transformations of the non-linear realisation,  is given by 
$$
2{\cal E}_\alpha {}_+ ^a= \nabla_\alpha x^a - \varepsilon^{ab}\nabla_\alpha y_b=0=\nabla_\alpha x^{a'}=\nabla_\alpha  y^{a'}
\eqno(3.21)$$ 
\par
At first sight the equations (3.21) do not look like the equations for the motion of the string. However, by multiplying by the matrix $s_\alpha{}^a\equiv \nabla _\alpha x^a$, and its inverse, 
we can write the first equation in  (3.21)  as 
$$
\sqrt {{-\gamma}} \gamma^{\alpha \beta} \nabla _\beta x^a =- \epsilon^{\alpha\beta} \nabla _\beta y^a \quad {\rm or \ equivalently } \ 
\sqrt { {-\gamma}} \gamma^{\alpha \beta} \nabla _\beta y^a = -\epsilon^{\alpha\beta} \nabla _\beta x^a 
\eqno(3.22)$$
where 
$$
\gamma _{\alpha\beta}= (s\eta s^T)_{\alpha\beta}\equiv \nabla _\alpha x^a \eta_{ab} \nabla _\beta x^b \quad {\rm and } \quad \gamma= det \gamma_{\alpha\beta}
\eqno(3.23)$$
\par
As a result equation (3.21) implies that  
$$
\sqrt {{-\gamma}} \gamma^{\alpha \beta} \nabla _\beta x^{\underline a }= -\epsilon^{\alpha\beta} \nabla _\beta y^{\underline a} \quad {\rm or \ equivalently } \ 
\sqrt { {-\gamma}} \gamma^{\alpha \beta} \nabla _\beta y^{\underline a} = -\epsilon^{\alpha\beta} \nabla _\beta x^{\underline a} 
\eqno(3.24)$$
\par
These are still not obviously the equations of motion of the string as we have the derivatives $\nabla_\alpha x^{\underline a }$ rather than 
$\partial_\alpha x^{\underline a }$. However, we note that we can write 
$$
\gamma _{\alpha\beta}=  \nabla _\alpha x^a \eta_{ab} \nabla _\beta x^b
= \nabla _\alpha x^a \eta_{ab} \nabla _\beta x^b+\nabla _\alpha x^{a^\prime} \eta_{a^\prime b^\prime} \nabla _\beta x^{b^\prime}= 
\nabla _\alpha x^{\underline a} \eta_{\underline a \underline b} \nabla _\beta x^{\underline b}= \partial _\alpha x^{\underline a} \eta_{\underline a \underline b} \partial_\beta x^{\underline b}
\eqno(3.25)$$
The last line follows from the fact that the line before it is SO(10,10) invariant and so the fields $\varphi^{ab^\prime}$, associated with the generators  $J_{ab^\prime}$ will be absent. Put another way we could carry out a $J_{ab^\prime}$ transformation to remove these fields and the object would be  unchanged. 
\par
The same argument can be used on the other covariant derivatives that appear in equation (3.24) as these equations are obviously  invariant under SO(10,10) transformations and so we can also  remove all the $\varphi ^{ab^\prime}$ terms by such a transformation. As a result we find the equations 
$$
\sqrt {{-\gamma}} \gamma^{\alpha \beta} \partial _\beta x^{\underline a} =- \epsilon^{\alpha\beta} \partial _\beta y^{\underline a} \quad {\rm or \ equivalently } \ 
\sqrt { {-\gamma}} \gamma^{\alpha \beta} \partial  _\beta y^{\underline a} = -\epsilon^{\alpha\beta} \partial _\beta x^{\underline a} 
\eqno(3.26)$$
Taking the derivative of the first equation we find the well known equation for the motion of the string. The brane dynamics in the background including the  level zero the fields, which are just the massless fields of the NS-NS sector of the IIA string,  can be found by making the replacement of equation (2.16) using the vielbein of reference [19]. 
\par
The equations of motion (3.20) are a mixture of traditional equations and inverse Higgs conditions [25]. The latter are conditions that allow one to algebraically solve for some of the fields of the non-linear realisation in terms of some of the other fields. 
We have only one  $I_c(SO(D,D))$  field in the non-linear realisation, namely    $\varphi_{ab^\prime}$. As we will see below,  the condition  
$\sqrt {{-\gamma}} \gamma^{\alpha \beta} \nabla _\beta x^{a^\prime} + \epsilon^{\alpha\beta} \nabla _\beta y^{a^\prime }=0 $ 
allows us to  solve for the combination 
$\sqrt {{-\gamma}} \gamma^{\alpha \beta} \partial _\beta x^{a^\prime} + \epsilon^{\alpha\beta} \partial _\beta y^{a^\prime }=0 $ 
in terms of $\varphi_{ab^\prime}$. Once this has been solved there are no  $I_c(SO(D,D)$  fields   in the equations which are just a function of 
$x^{\underline a}$ and $y^{\underline b}$ which just transform under the rigid transformations of equation (2.12). 
\par
An unexpected feature of the above calculation  is the choice of local subalgebra of equation (3.8) rather than the naively expected subalgebra $SO(1,1) \otimes SO(1,1) \otimes SO(D-2) \otimes SO(D-2) $. In the latter case  one would have two fields of $\varphi$ type corresponding to the generators $J_{ab^\prime}$ and $S_{ab^\prime}$, 
rather than one corresponding to just $J_{ab^\prime}$. As we mentioned already  the calculation of reference [16] is rather immune to the choice of subalgebra as it works with the quantity $M$, mentioned above, that is invariant under local transformations. 
\par
In view of the unusual way the brane dynamics appears in the non-linear realisation when one uses the Cartan forms  it is instructive to examine in detail what happens at the linearised level. Using equations (2.15) and (3.6) we find, up to terms that are at most  linear in the fields $\varphi$,  that the Cartan forms of equation (3.16) are given by  
$$
\nabla_{\alpha}x^{\underline{a}} = \partial_{\alpha}x^{\underline{a}} + 2\partial_{\alpha}x^{\underline{c}}\phi^{(1)}{}_{\underline{c}}{}^{\underline{a}} + 2\partial_{\alpha}y^{\underline{c}}\phi^{(2)}{}_{\underline{c}}{}^{\underline{a}}
 + \ldots 
\eqno(3.27)$$
$$
\nabla_{\alpha}y^{\underline{a}} = \partial_{\alpha}y^{\underline{a}} + 2\partial_{\alpha}y^{\underline{c}}\phi^{(1)}{}_{\underline{c}}{}^{\underline{a}} + 2\partial_{\alpha}x^{\underline{c}}\phi^{(2)}{}_{\underline{c}}{}^{\underline{a}}
 + \ldots 
\eqno(3.28)$$
 if  we were to take the local group element $g_h$ to be of that most general form, that is, of the form, namely   $g_h= 1+  (\phi^{(1)}{}_{\underline{a} \underline{b}} J^{\underline{a}\underline{b}} + \phi^{(2)}{}_{\underline{a} \underline{b}} S^{\underline{a}\underline{b}})$. This would be the case if we had not used the local subalgebra to set some of the fields in $g_h$ to zero, whereupon  the  actual group element is of the form $g_h=e^{-\varphi^{ab^\prime} J_{ab^\prime}}$ as given  in equation (3.15).   To make the content of  equations (3.27) and (3.28) apparent  we choose static gauge, that is, 
$\partial_{\alpha}x^c = \delta_{\alpha}^c$. Examining the first equation of motion of (3.21), that is $\nabla_\alpha x^a - \varepsilon^{ab}\nabla_\alpha y_b=0$, we find that at zeroth order we must take 
$$
\partial_{\alpha}x^c = \delta_{\alpha}^c\  \quad {\rm and } \quad 
\partial_{\alpha}y_a= -\epsilon_{\alpha a}
\eqno(3.29)$$
\par
Taking $\underline a= a^\prime$ in equations (3.23) and (3.24) we find that then  at the lowest  non-trivial level  they become  
$$\nabla_{\alpha}x^{a'} = \partial_{\alpha}x^{a'} + 2\phi^{(1)}{}_\alpha{}^{a'} - 2\epsilon_{\alpha c}\phi^{(2)}{}^{ca'} +\ldots 
\eqno(3.30)$$
and
$$
\nabla_{\alpha}y^{a'} = \partial_{\alpha} y^{a'} - 2\epsilon_{\alpha c}\phi^{(1)}{}^{ca'} + 2\phi^{(2)}{}_{\alpha}{}^{a'} +\ldots 
\eqno(3.31)$$ 
 where  we have taken the general group element $g_h$ but we with  only   $\phi^{(1)}{}^{a'}{}_c$ and $\phi^{(2)}{}^{a'}{}_c$ to be  non-zero. We may rewrite  equations (3.30) and (3.31) as 
$$
\nabla^\alpha x^{a^\prime} - \epsilon^{\alpha \delta}\nabla_\delta y^{a^\prime}= \partial^\alpha x^{a^\prime} - \epsilon^{\alpha \delta}\partial_\delta y^{a^\prime} +4(\phi^{(1)} {}^{\alpha a^\prime} -\epsilon^{\alpha \delta }\phi^{(2)}{}_\delta{}^{a^\prime})+\ldots 
\eqno(3.32)$$
and 
$$
\nabla_\alpha x^{a^\prime} + \epsilon^{\alpha \delta}\nabla_\delta y^{a^\prime}= 0+\ldots 
\eqno(3.33)$$
We observe that only the combination $\phi^{(1)} {}^{\alpha a^\prime} -\epsilon^{\alpha \delta }\phi^{(2)}{}_\delta{}^{a^\prime}$ occurs in the equations of motion and that    if we set the right-hand side of equation (3.32) to zero this quantity  will be solved in terms of $\partial_\alpha x^{a^\prime} - \epsilon^{\alpha \delta}\partial_\delta y^{a^\prime}$. This is an example of the so called inverse Higgs effect. The orthogonal combination, that is $\phi^{(1)} {}^{\alpha a^\prime} 
+\epsilon^{\alpha \delta }\phi^{(2)}{}_\delta{}^{a^\prime}$,  does not appear in the equations of motion (3.21) and so we cannot solve for this combination in terms of any of the fields $\partial _\alpha x^{\underline a}$. Our choice of local subalgebra ${\cal H}$ of equation (3.8) and in particular the inclusion of  the generators $L_{ab^\prime}$ ensures that
the orthogonal combination $\phi^{(1)} {}^{\alpha a^\prime} 
+\epsilon^{\alpha \delta }\phi^{(2)}{}_\delta{}^{a^\prime}$ arises in the local subgroup and not in the group element $g_h$ and so is automatically absent from the theory. This observation can be used to justify the choice of local subgroup. As a result  the combination in equation (3.32) is proportional to $\varphi ^{\alpha a^\prime}$. 
\par
Thus we see that setting $\nabla_\alpha x^{a'}=0=\nabla_\alpha  y^{a'}$, as we did in equation (3.21), has the effect of solving for the $\varphi$ fields that arise in the group element $g_h$ and also enforcing the equation 
$$
\nabla_\alpha x^{a^\prime} + \epsilon^{\alpha \delta}\nabla_\delta y^{a^\prime}+\ldots =0
\eqno(3.34)$$
which we recognise as the linearised version of the equation of motion of  equation (3.26) in static gauge. 
We will find that the above pattern that emerges for the string case also occurs  for all the other branes we study in this paper. 
\medskip
{\bf 4. Branes in eleven dimensions }
\medskip
In this section we will consider  the dynamics of the M2 and the  M5 branes. 
\medskip
{\bf 4.1 The M2 brane}
\medskip
In this section we will construct the dynamics of the M2 brane in eleven dimensions using the non-linear realisation $E_{11}\otimes_s l_1$ as explained in section two. The dynamics of the M2 brane was  found in 
the classic paper of reference [26] and a formulation involving the dual fields $x^a$ and $x^{a_1a_2}$ was given in reference [25]. A partial account of the dynamics of the M2 brane from the non-linear realisation $E_{11}\otimes_s l_1$ was given in reference [9]. 
\par 
The eleven dimensional theory emerges when  we take the decomposition of $E_{11}$ into SL(11) which appears when we delete node eleven of the $E_{11}$ Dynkin diagram. We will first constructing the brane in the absence of background fields and so we consider the non-linear realisation of the algebra  $I_c(E_{11})\otimes_s l_1$ which in the decomposition to SL(11) has the generators 
$$
I_c(E_{11})=\{J_{\underline a_1\underline a_2}, \ S_{\underline a_1 \underline a_2\underline a_3} , \ S_{\underline a_1, \ldots , \underline a_6} ,\ 
S_{\underline a_1, \ldots \underline a_8,\underline b} , \ldots \}
\eqno(4.1.1)$$
where $\underline a_1, \underline a_2\ldots =0,1,\ldots 10$ while   
the  generators  of the vector representations are  
$$
l_1=\{P_{\underline a} , \ Z^{\underline a_1\underline a_2}, \ Z^{\underline a_1\ldots \underline a_5}, \  Z^{\underline a_1\ldots \underline a_8},  Z^{\underline a_1\ldots \underline a_7,\underline b}, \ldots  \}
\eqno(4.1.2)$$
The algebra the generators of equation (4.1.1) and (4.1.2)  obey can be found in reference [28]. 
\par
We choose the local subalgebra ${\cal H}$ to be given by 
$$
{\cal H}= \{J_{a_1a_2},\  J_{a_1^\prime a_2^\prime }\ , \hat S\equiv \epsilon^{a_1a_2a_3} S_{a_1a_2a_3}\ , L_{ab^\prime}\equiv 2J_{ab^\prime}+\epsilon _{ae_1e_2} S^{e_1e_2}{}_{b^\prime}, S_{ab_1^\prime b_2^\prime}\ ,  $$
$$
\hat S_{a_1^\prime a_2^\prime a_3^\prime}\equiv  S_{a_1^\prime a_2^\prime a_3^\prime} +{1\over 3} \epsilon ^{e_1e_2e_3} S_{e_1e_2e_3 a_1^\prime a_2^\prime a_3^\prime}\  ,\ S_{ b^\prime_1 \ldots b_6^\prime} ,\ ,\ 
S_{a_1 b^\prime_1 \ldots b_5^\prime} ,\ ,\ 
S_{a_1a_2 b^\prime_1 \ldots b_4^\prime} ,\ \ldots \}
\eqno(4.1.3)$$
where $a_1,  a_2, \ldots =0,1,2$ and $a_1^\prime a_2^\prime, \ldots =3,4\ldots ,10$. A motivation for this choice of local subalgebra will be given when we analyse the linearised theory at the end of this section.  
\par
As discussed in section two the non-linear realisation is constructed from the group element $g=g_lg_h$. The group element $g_l$ is given by 
$$
g_l = e^{x^{\underline a} P_{\underline a}} e^{x_{\underline a_1\underline a_2}Z^{\underline a_1\underline a_2}} e^{x_{\underline a_1\ldots \underline a_5}Z^{\underline a_1\ldots \underline a_5}}\ldots 
\eqno(4.1.4)$$
while the group element $g_h= e^{\varphi \cdot S}$ can be chosen to take the   form 
$$
g_h= e^{-\varphi _a{}^{b^\prime} J^a{}_{b^\prime}}\ldots 
\eqno(4.1.5)$$
We note that  the  generators which are in $I_c(E_{11})$ but not in ${\cal H}$ can be chosen to be $J^a{}_{b^\prime}$ up to generators of level two. 
\par 
The algebra obeyed by the generators of  ${\cal H}$ of equation (4.3), up to commutators that involve level two generators, is given   
$$
[J_{ab},  L_{cd^\prime } ]=-\eta _{ac} L_{bd^\prime}+\eta _{bc} L_{ad^\prime}
 ,\quad 
[ L_{ab^\prime }, L_{cd^\prime} ]= 0, \quad
[ L_{ab^\prime }, \hat S]= -6 L_{ab^\prime } , \quad 
$$
$$
[ L_{ab^\prime }, S_{d c_1^\prime c_2^\prime} ]= -2\eta_{ad} \hat S_{b^\prime c_1^\prime c_2^\prime } +2 \epsilon _{a de}\eta_{b [c_1^\prime } L^e{}_{ c_2^\prime]} , \ 
[ L_{ab^\prime }, \hat S_{c_1^\prime c_2^\prime c_3^\prime} ]= 
6\eta_{b^\prime [ c_1^\prime} S_{|a| c_2^\prime c_3^\prime]} ,\ 
$$
$$
[S^{a b_1^\prime b_2^\prime} , \hat S_{c_1^\prime c_2^\prime c_3^\prime} ]=
2S^{a b_1^\prime b_2^\prime }{}_{c_1^\prime c_2^\prime c_3^\prime}
-3\delta^{b_1^\prime b_2^\prime}_{[c_1^\prime c_2^\prime} L^a{}_{c_3^\prime]}
$$
$$
[\hat S , \hat S _{a_1^\prime a_2^\prime a_3^\prime} ]= 6 \hat S_{a_1^\prime a_2^\prime a_3^\prime} , \ 
[\hat S ,  S _{a b_1^\prime b_2} ]= 0 , \ \ [ S_{ a_1^\prime a_2^\prime a_3^\prime} , S_{b_1^\prime b_2^\prime b_3^\prime} ]=2S_{a_1^\prime a_2^\prime a_3^\prime b_1^\prime b_2^\prime b_3^\prime}
$$
$$
[S^{a c_1^\prime c_2^\prime} , S_{b d_1^\prime d_2^\prime}]=
2S^{a c_1^\prime c_2^\prime}{}_{b d_1^\prime d_2^\prime}
-2 \delta ^{c_1^\prime c_2^\prime}_{d_1^\prime d_2^\prime} J^a{}_{b}
-8 \delta^{a [ c_1^\prime }_{b [ d_1^\prime} J^{c_2^\prime]} {}_{d_2^\prime]}
\eqno(4.1.6)$$
The commutators involving  $J_{ab}$ and $J_{a^\prime b^\prime}$ are as one naively expects and as a result  we have not written them down. 
\par
Rather than work with the generators  of the $l_1$ representation as given in equation (4.1.2) it will be advantageous to instead use the following objects
$$
 P_{a^\prime} ,\  N^\pm_a\ ,\ Z_{a_1a_2}\ ,\ Z_{ab^\prime} 
,\ 
N^\pm_{a_1^\prime a_2^\prime},\ Z_{a_1a_2b_1^\prime b_2^\prime b_3^\prime } ,\  Z _{a b_1^\prime \ldots  b_4^\prime },\  \hat Z _{ a_1^\prime \ldots  a_5^\prime },\ ,\  Z _{ a_1 \ldots  a_5 }, \ldots  
\eqno(4.1.7)$$
where  
$$
 N^\pm_a\equiv P_a\pm {1\over 2} \epsilon_{ae_1e_2}Z^{e_1e_2} ,\ 
N^\pm _{a_1^\prime a_2^\prime}= Z_{a_1^\prime a_2^\prime}\mp {1\over 3!}\epsilon ^{e_1e_2e_3} Z_{e_1e_2e_3 a_1^\prime a_2^\prime}, \ 
$$
$$
 \hat Z _{ a_1^\prime \ldots  a_5^\prime }= Z_{ a_1^\prime \ldots  a_5^\prime }+{1\over 3!}\epsilon ^{e_1e_2e_3} ({5\over 3} Z_{e_1e_2e_3 [a_1^\prime \ldots a_4^\prime , a_5^\prime ]}- Z_{e_1e_2e_3 a_1^\prime \ldots a_5^\prime }), 
$$
The generators at level four in the vector representation  have eight Lorentz indices and we include such a  contribution in the last equation as it was useful in the higher level calculations not presented in this paper. 
\par
The generators of the the vector representation given in   equation (4.1.7) have the following   commutators with the generators of ${\cal H}$:  
$$ 
[L_{ab'}, P_{c'}] = 2\eta_{b'c'}N^+_a , \ \
[L_{ab'}, N_c^+] = 0, \ \
[L_{ab'}, N_c^-] = -4\eta_{ac}P_{b'} + 4\varepsilon_{acd}Z^{db'}, \ \
$$
$$
[L_{ab'}, N_{c'_1c'_2}^-] = 0 , \ \ [L_{ab'}, N_{c'_1c'_2}^+] =-8 \eta_{b^\prime [c_1^\prime} Z_{c_2^\prime] a} + 2\epsilon _{a }{}^{e_1 e_2} Z_{e_1e_2 b^\prime c'_1c'_2}$$
$$
[L_{ab'}, Z_{cd'}] = 2\eta_{b'd'}\varepsilon_{ace}N_e^+ - 2\eta_{ac}N^-_{b'd'} , \ \ 
$$
$$ [\hat{S}, N_c^{\pm}] = 6N_c^{\pm} , \ \
[\hat{S}, N_{c'_1c'_2}^\pm] =6  N_{c'_1c'_2}^\pm , \ \
[\hat{S}, Z_{ab'}] = 0, \ \
[\hat{S}, Z_{c_1c_2d'_1d'_2d'_3}] = 0 ,\ \ 
$$
$$ [S_{ab'_1b'_2}, P_{c'}] = 2\eta_{c'[b'_1}Z_{b'_2]a} , \ \
[S_{ab'_1b'_2}, N_c^\pm] = \eta_{ac}N_{b'_1b'_2}^\mp , \ \ 
$$
$$
[S_{ab'_1b'_2}, N_{c'_1c'_2}^{\mp}] = -2\delta_{c'_1c'_2}^{b'_1b'_2}N^{\pm}_a + Z^{\pm}_{ab'_1b'_2c'_1c'_2} , \ \
[S_{ab'_1b'_2}, Z_{cd'}] =Z_{acb'_1b'_2d'}-4\eta_{d^\prime[b'_1} P_{b'_2]} , $$
$$ 
[\hat{S}_{a'_1a'_2a'_3}, P_{c'}] = 3\delta^{c'}_{[a'_1}N_{a'_2a'_3]}^- , \ \
[\hat{S}_{a'_1a'_2a'_3}, N_c^+] = 0 , \ \
[\hat{S}_{a'_1a'_2a'_3}, N_c^-] = -\varepsilon^{ce_1e_2}Z_{e_1e_2a'_1a'_2a'_3} , \ \
$$
$$
[\hat{S}_{a'_1a'_2a'_3}, N_{c'_1c'_2}^-] =\hat Z_{a'_1a'_2a'_3c'_1c'_2} , \ \ [\hat{S}_{a'_1a'_2a'_3}, Z_{bc'}] = Z_{ba'_1a'_2a'_3c'}^+ , 
$$
$$
[\hat{S}_{a'_1a'_2a'_3}, N_{c'_1c'_2}^+] =-12 \delta_{c_1^\prime c_2^\prime }^{ [ a_1^\prime a_2^\prime }P^{a_3^\prime ]}+\hat Z_{a'_1a'_2a'_3c'_1c'_2}+\ldots 
\eqno(4.1.8)$$
where $+\ldots$ means higher level generators. We observe that the generators 
$$
N_a^+ , \ N_{a_1^\prime a_2 ^\prime }^- , Z_{a_1^\prime \ldots a_5^\prime}, \ldots $$ 
 form an irreducible representation of the local subgroup ${\cal H}$ up to the level we have calculated. 
\par 
The dynamics is just a set of equations that are invariant under the transformations of equation (2.15). The Cartan form ${\cal V}_l$ can be expressed as 
$$
{\cal V}_l=  (\nabla x^{\underline a} P_{\underline a}+  \nabla x_{\underline a_1\underline a_2} Z^{\underline a_1\underline a_2}+  \nabla x _{\underline a_1\ldots \underline a_5} Z^{\underline a_1\ldots \underline a_5}
+  \ldots  )
\eqno(4.1.9)$$
where where  $\nabla\equiv d\xi^\alpha \nabla_\alpha$ and we will write the Cartan forms as forms. When written  in terms of the generators of equation (4.1.7) The Cartan form ${\cal V}_l$ takes the form 
$$ {\cal{V}} = {\cal E}_a^+N_+^a + {\cal E}_a^-N_-^a + {\cal E}^{a^\prime} P_{a'} + {\cal E}_{a_1'a_2'}^-N^-_{a'_1a'_2} + {\cal E}^+_{a'_1a'_2}N^+_{a'_1a'_2} + {\cal E}_{ab'}Z^{ab'} +\ldots 
\eqno(4.1.10) $$

$$
{\cal E}^{a^\prime}\equiv \nabla x^{a'}\ ,\ {\cal E}_a^\pm \equiv {1\over 2} 
(\nabla  x^{ a}\mp \epsilon ^{a e_1e_2} \nabla x_{ e_1 e_2} ) \ ,\ 
$$
$$
{\cal E}_{a_1'a_2'}^\pm \equiv 
 {1\over 2} (\nabla x_{ a_1^\prime a_2^\prime}\pm 
\epsilon ^{ e_1e_2e_3} \nabla x_{ e_1e_2e_3 a_1^\prime a_2^\prime})
\ ,\ 
{\cal E}_{ab'}\equiv 
\nabla x_{ ab^\prime} ,
\eqno(4.1.11)$$
\par
Under the transformation $h=1-(\Lambda S+ \Lambda^{ab^\prime} L_{ab^\prime} + \Lambda ^{ab_1^\prime b_2^\prime}S_{ab_1^\prime b_2^\prime}+ \Lambda ^{b_1^\prime b_2^\prime  b_3^\prime}S_{b_1^\prime b_2^\prime  b_3^\prime} )$ we find, using equation (2.15), that the Cartan forms transform as 
$$
\delta{\cal E}_a^- = -2\Lambda^{ab'_1b'_2}{\cal E}^+_{b'_1b'_2} + 6 \Lambda{\cal E}^-_a$$
$$ \delta{\cal E}_a^+ = {\cal E}^{b'}\Lambda^{ab'} + 2{\cal E}^{c}{}_{b'}{\epsilon}_{dca}\Lambda^{db'} - 2\Lambda_{a}{}^{b'_1b'_2}{\cal E}^-_{b'_1b'_2} + 6\Lambda{\cal E}_a^+ $$
$$ \delta({\cal E}^{a'}) = -4{\cal E}^{b-}\Lambda_{b }{}^{a^\prime} - 12\Lambda^{c'_1c'_2a'}{\cal E}^+_{c'_1c'_2} $$
$$ \delta({\cal E}^{ab'}) = 4{\epsilon}_{dc}{}^{a}{\cal E}^{c-}\Lambda^{db'} + 8{\cal E}^+{}^{c'b'}\Lambda^{a}{}_{c'} - 2\Lambda^{ad'b'}{\cal E}_{d'} $$
$$\delta{\cal E}^+_{a'_1a'_2} = \Lambda^{ca'_1a'_2}{\cal E}_c^- + 6\Lambda{\cal E}^+_{a'_1a'_2}$$
$$ \delta{\cal E}^-_{a'_1a'_2} = -2{\cal E}_{e[a'_2|}\Lambda^e{}_{|a'_1]} + \Lambda^{c}{}_{a'_1a'_2}{\cal E}_c^+ +3\Lambda_{c'a'_1a'_2}\nabla_{\alpha}x^c + 6\Lambda{\cal E}_{a'_1a'_2}^- 
\eqno(4.1.12)$$
\par
Examining the variations of the Cartan form of equation (4.1.12) one sees that one can preserve all the symmetries of the non-linear realisation by  setting  
$$ 
\delta {\cal E}_a^- = 0 = {\cal E}^{a'} = {\cal E}^+_{a'_1a'_2} = {\cal E}_{ab'} 
\eqno(4.1.13)$$
which are equivalent  to the equations  
$$
\nabla_\alpha x^{ a}+ \epsilon ^{a e_1e_2} \nabla_\alpha x_{ e_1 e_2} =0 , \ {\rm or \ equivalently}  ,\ 
\nabla_\alpha x_{ a_1 a_2}-{1\over 2}\epsilon_{a_1a_2 c} \nabla_\alpha x^{ c}=0
\eqno(4.1.14)$$  
 $$
\nabla_\alpha x^{ a^\prime} =  \nabla_\alpha x_{ a b^\prime } =0 
\eqno(4.1.15)$$
$$
\nabla_\alpha x_{ a_1^\prime a_2^\prime}+ 
\epsilon ^{ e_1e_2e_3} \nabla_\alpha x_{ e_1e_2e_3 a_1^\prime a_2^\prime}= 0 
\eqno(4.1.16)$$
\par
At first sight these are not the familiar equations of motion for the M2 brane. However, as discussed later  in section seven  in the context of the dynamics of general  branes, we may first write  equations (4.14) as 
$$
\sqrt {-\gamma} \gamma ^{\alpha \beta} \nabla _\beta x^a=  
\epsilon ^{\alpha \beta \gamma }\nabla _\beta 
x^{a b  } \nabla _{\gamma} x_{b} \ ,
\eqno(4.1.17)$$
or equivalently,  
$$
\sqrt {-\gamma} \gamma ^{\alpha \beta} \nabla _\beta x^{a_1 a_2} =
{1\over 2} \epsilon ^{\alpha \gamma_1  \gamma _{2} }
\nabla _{\gamma_1} x_{b_{1}} \nabla _{\gamma_2} x_{b_{2}}
\eqno(4.1.18)$$
where 
$$
\gamma _{\alpha \beta} \equiv  (s\eta s^T)= \nabla_\alpha x^a \eta_{ab} \nabla _\beta x^b \quad {\rm and }\quad \gamma= det \gamma _{\alpha \beta} 
\eqno(4.1.19)$$
As a result and using equations (4.1.15) and (4.1.16) we may write the dynamical equations as 
$$
\sqrt {-\gamma} \gamma ^{\alpha \beta} \nabla _\beta x^{\underline a}=  
\epsilon ^{\alpha \beta \gamma }\nabla _\beta 
x^{\underline a \underline b  } \nabla _{\gamma} x_{\underline b}\eqno(4.1.20)$$
Taking into account that the only $\varphi$ fields are those associated with the Lorentz rotations $J_{ab^\prime}$ and  using arguments similar to those at the end of section three we conclude that we may write the equations of motions as  
$$
\sqrt {-\gamma} \gamma ^{\alpha \beta} \partial _\beta x^{\underline a}=  
\epsilon ^{\alpha \beta \gamma }\partial_\beta 
x^{\underline a \underline b  } \partial _{\gamma} x_{\underline b}\eqno(4.1.20)$$
Acting on this equation with a derivative we recover the well known equation for the motion of the M2  brane. It would be interesting to carry out the analysis of the brane dynamics at higher levels in the coordinates and in particular to systematically incorporate the level three coordinates. In particular it would be useful to know if there is non-trivial information encoded in the higher level coordinates that is not contained in the equations for the lower level coordinates discussed above. 
\par
The brane in the presence of the background fields, including those of eleven dimensional supergravity,  can be found  by making the replacement of equation (2.1.16) using the veilbein given in reference [27]. 
\par 
It is instructive to carry out the linearised analysis of the equations of motion  (4.1.14) and (4.1.15) in order to understand their content in more detail.  We begin by computing the Cartan forms when taking the group element $g_h$ to have the most general form and so  given by 
$$
g_h = e^{\phi _{\underline a \underline b} J^{\underline a \underline b} + \phi_{\underline a_1 \underline a_2 \underline a_3}S^{\underline a_1 \underline a_2 \underline a_3} + \phi_{\underline a_1 \ldots \underline a_6} S^{\underline a_1 \ldots \underline a_6+\ldots }}
\eqno(4.1.21)$$
We can use  the  local subgroup ${\cal H}$ to restrict the group element as indeed we have done above in  arriving  at the  group element $g_h$ of equation (4.1.5). We can recover the later by restricting the fields $\phi$. However, it will be useful  to consider a  general group element $g_h$ so we can  motivate the  choice of  the local subgroup ${\cal H}$. Using equation (2.15) we find that 
$$
\nabla_\alpha x^{\underline c} = \partial_\alpha  x^{\underline c}+6\phi^{\underline e_1 \underline e_2 \underline c}\partial _\alpha x_{\underline e_1 \underline e_2}+12.30 \phi^{\underline e_1 \ldots \underline e_5 \underline c}\partial _\alpha x_{\underline e_1 \ldots \underline e_5}+2\partial_\alpha x^{\underline e}\phi_{\underline e}{}^{ \underline c}+\ldots
\eqno(4.1.22)$$
$$
\nabla_\alpha x^{\underline c_1\underline c_2} = \partial_\alpha x^{\underline c_1\underline c_2}-3 \phi^{\underline c_1 \underline c_2 \underline e} \partial_\alpha x_{\underline e}  +{5!\over 2} \phi_{\underline e_1 \underline e_2 \underline e_3 }\partial_\alpha x^{\underline e_1 \underline e_2 \underline e_3 \underline c_1 \underline c_2} +4\partial_\alpha x^{\underline e [\underline c_2 |} \phi _{\underline e}{}^{|\underline c_1 ]}+\ldots 
\eqno(4.1.23)$$
$$
\nabla_\alpha x^{\underline c_1 \ldots \underline c_5} = 
\partial_\alpha x^{\underline c_1 \ldots \underline c_5} 
-\phi^{ [\underline c_1 \underline c_2 \underline c_3 }\partial_\alpha x^{\underline c_4 \underline c_5 ]}  -3\phi^{\underline c_1 \ldots \underline c_5 \underline e} \partial_\alpha x_{\underline e} 
+10 \partial _\alpha x^{\underline e [ \underline  c_2 \ldots \underline c_5} \phi_{\underline e }{}^{\underline c_1 ]} +\ldots 
\eqno(4.1.24)$$
if we only keep terms which are at most linear in the $\phi$ fields. 

If we adopt static gauge $\partial_\alpha x^c=\delta_\alpha^c$,   then equation of motion (4.14)  implies that $\partial_\alpha x_{a_1a_2}= {1\over 2} \epsilon_{\alpha  a_1a_2}$ at lowest order. Keeping terms that are at most linear in either the $x$ or $\phi$ fields,  we find that the Cartan forms are given by 
$$
\nabla _\alpha x^{ c^\prime} - 
\epsilon _{\alpha }{}^{\beta e} \nabla _\beta x_{e}{}^{c^\prime}= 
\partial _\alpha x^{ c^\prime} - 
\epsilon _{\alpha }{}^{\beta e} \partial _\beta x_{e}{}^{c^\prime}+\ldots 
\eqno(4.1.25)$$
while 
$$
\nabla _\alpha x^{ c^\prime} + 
\epsilon _{\alpha }{}^{\beta e} \nabla _\beta x_{e}{}^{c^\prime}=
\partial _\alpha x^{ c^\prime} +
\epsilon _{\alpha }{}^{\beta e} \partial _\beta x_{e}{}^{c^\prime}
+2 (2 \phi_\alpha{}^{c^\prime} +3 \epsilon _{\alpha e_1e_2} \phi^{e_1e_2 c^\prime} )+\ldots 
\eqno(4.1.26)$$
\par
The equations of motion (4.1.15) set the expressions in equations (4.1.25) and (4.1.26) to zero. We observe that they only contain the combination 
$2 \phi_\alpha{}^{c^\prime} +3 \epsilon _{\alpha e_1e_2} \phi^{e_1e_2 c^\prime} $ and not the orthogonal expression 
$3 \phi_\alpha{}^{c^\prime} -2 \epsilon _{\alpha e_1e_2} \phi^{e_1e_2 c^\prime} $. Ensuring that the latter combination is not contained in the group element $g_h$ motivates our choice of local subalgebra of equation (4.1.3). One finds that the former combination is proportional to $\varphi_\alpha{}^{c^\prime} $. Setting to zero equation (4.1.26) enables us to solve for the expression involving $x$'s in terms of the $\phi$ fields and it is an example of the inverse Higgs effect. Setting to zero equation (4.1.25) is the correct linearised equation of motion for the M2 branes and it agrees with equation (4.20) at the linearised level. 
\medskip
{\bf 4.2 The M5 brane}
\medskip
As the five brane has a six dimensional world volume we take the indices to have the range $a,b,\ldots = 0,1,\ldots , 5$ and $a^\prime,b^\prime,\ldots = 6,\ldots ,10$. The local subgroup ${\cal H}$ is a subgroup of $I_c(E_{11})$ and we choose it to contain the generators   
$$
{\cal H}= \{J_{ab}, \ J_{a^\prime b^\prime} ,\  L_{ab^\prime} \equiv J_{ab^\prime}+{2\over 5!} \epsilon_{a e_1\ldots e_5} S^{e_1\ldots e_5}{}_{b^\prime}, \ 
S_{-a_1a_2a_3},\  S\equiv {1\over 6!}\epsilon^{a_1\ldots a_6}S_{a_1\ldots a_6}, \ldots \}
\eqno(4.2.1)$$
where we adopt the notation 
$$
X_{\pm a_1a_2a_3}= {1\over 2} (X_{-a_1a_2a_3}\pm{1\over 3!} \epsilon _{a_1a_2a_3 b_1b_2b_3} X^{b_1b_2b_3})
\eqno(4.2.2)$$
for any object $X_{a_1a_2a_3}$.
These generators obey the algebra 
$$
[L_{ac^\prime} , L_{bd^\prime} ]= 0  , \ [ S_{-a_1a_2a_3} , S_{-b_1b_2b_3} ]=0, \ [S, S_{-a_1a_2a_3} ]= {1\over 2} S_{-a_1a_2a_3}, \ 
[S , L_{ab^\prime} ]= {1\over 2} L_{ab^\prime}, \ldots 
\eqno(4.2.3)$$\
as well as the expected commutators with $J_{ab}$ and  $J_{a^\prime b^\prime}$. 
\par
In the last section we took  a preferred basis of the generators of the vector representation, and so the Cartan forms, however, here we will take a more direct approach  and compute the variations of the Cartan form as they appear in equation (4.1.9) under the local subgroup transformations  using equation (2.15). 
One finds that 
$$
\delta (\nabla   x^a)= 6\Lambda_{+}{}^{e_1e_2 a} \nabla   x_{e_1e_2} + {1\over 2}  \epsilon  ^{ae_1\ldots e_5 } \Lambda \nabla   x_{e_1\ldots e_5} +2 \nabla  x^{c^\prime} \Lambda_{c^\prime} {}^a , 
$$
$$
\delta (\nabla   x^{a^\prime})= 2(\nabla  x^c+\epsilon^{ce_1\ldots e_5} \nabla   x_{e_1\ldots e_5})
\Lambda_c{}^{a^\prime} 
$$
$$
\delta (\nabla   x_{a_1a_2})=-3 \Lambda_{+a_1a_2 e } \nabla  x^e +{5!\over 2} \Lambda_{+e_1e_2e_3} \nabla  x^{e_1e_2e_3 a_1a_2}+\ldots 
$$
$$
\delta (\nabla   x^{a_1\ldots a_5})=-\Lambda _+{}^{[a_1a_2a_3} \nabla   x^{a_4a_5]} +{1\over 2.5!} \epsilon ^{a_1\ldots a_5 e} \Lambda\nabla   x_{e}+{2\over 5!} \epsilon ^{a_1\ldots a_5 b} \Lambda_b{}_{c^\prime} \nabla  x^{c^\prime}+\ldots 
$$
$$
\delta (\nabla   x^{a_1\ldots a_4 b^\prime})=-{2\over 5!} 
\Lambda^{+[ a_1a_2a_3 } \nabla  x^{a_4 ] b^\prime}  
\eqno(4.2.4)$$
where $+\ldots $ refer to higher level terms which we are not considering here and $\nabla= d\xi^{\underline \alpha} \nabla_{\underline \alpha}$. One can also carry out a variation of the Cartan forms with a general group element $h$ and then impose the conditions 
$\Lambda_a{}^{c^\prime}= -{3\over 2} \epsilon_{a e_1\ldots e_5} \Lambda^{e_1\ldots e_5 c^\prime}$  as well as the self duality conditions on $\Lambda_{+ a_1a_2a_3}$ to ensure that the transformation is in the local subalgebra.  In particular we find that  under the above transformations of the local subalgebra  that 
$$
\delta {\cal E}^a =0 , \ \delta {\cal E}^{a^\prime}= 2{\cal E} ^b \Lambda _b{}^{a^\prime} ,\ \delta {\cal E}_{a_1a_2} = -3 \Lambda _{+a_1a_2 b}{\cal E}^b+\ldots 
\eqno(4.25)$$
where 
$$
{\cal E} {}^a\equiv \nabla x^a + \epsilon^{a b_1\ldots b_5} 
\nabla x_{b_1\ldots b_5} , \ 
{\cal E} {}^{a^\prime } = \nabla x^{a^\prime} , \ 
{\cal E}_{a_1a_2}^L\equiv \nabla x_{a_1 a_2 }
\eqno(4.2.6)$$
\par
We first consider the {\bf linearised } theory. From equation (4.2.6) we find that we can  adopt the conditions 
$$
{\cal E}_\alpha {}^a =0 , \ {\cal E}_\alpha {}^{a^\prime } =0 
\eqno(4.2.7)$$
as well as 
$$
{\cal E}^L _{[\alpha e_1e_2] }=0
\eqno(4.2.8)$$
and, up to the adoption of higher level constraints, they are invariant. 
We are working at the linearised level as the last equation antisymmetises indices whose transformation character is different.
 \par
To understand the meaning of the equations (4.2.7) and (4.2.8) for the linearised theory it  is instructive to calculate the Cartan forms at the linearised level. For this we take static gauge $\partial_\alpha x^a =\delta_\alpha ^a$, $\partial_\alpha x^{a_1\ldots a_5}={1\over 5!}\epsilon^{\alpha a_1\ldots a_5}$  and keep only terms that are linear in the $x$'s and the $\phi$'s. One finds that {\bf if} one takes the general group element of equation (4.1.21) and uses 
 equations (4.1.22) and (4.1.23) that  
$$
\nabla_\alpha x^{a^\prime}  -\epsilon_{\alpha}{}^{ \beta b_1\ldots b_4} \nabla _\beta x_{b_1\ldots b_4}{}^{ a^\prime }=\partial_\alpha x^{a^\prime}  -\epsilon^{\alpha \beta b_1\ldots b_4} \partial _\beta x_{b_1\ldots b_4}{}^{ a^\prime },\ 
\eqno(4.2.9)$$
$$
\nabla_\alpha x^{a^\prime}  +\epsilon_{\alpha}{}^{ \beta b_1\ldots b_4} \nabla _\beta x_{b_1\ldots b_4}{}^{ a^\prime }= 
\partial_\alpha x^{a^\prime}  +\epsilon^{\alpha \beta b_1\ldots b_4} \partial _\beta x_{b_1\ldots b_4}{}^{ a^\prime }+4\hat \phi_\alpha{}^{a^\prime}  ,
\eqno(4.2.10)$$
$$
\nabla_\alpha x_{a_1a_2} = \partial_\alpha x_{a_1a_2} -6\hat \phi_{-\alpha a_1a_2}
\eqno(4.2.11)$$
where
$$
\hat \phi_{\alpha}{}^{a^\prime}= {1\over 2} (\phi_{\alpha}{}^{a^\prime}
+{3\over 2} \epsilon_{\alpha e_1\ldots e_5 } \phi^{e_1\ldots e_5 a^\prime} ) 
\eqno(4.2.12)$$
We have chosen our local subalgebra just so that the $\phi$'s that occur in the Cartan forms contain the combination that occurs in equation (4.2.12) and not the orthogonal combination. As a result we take the generator $L_{ab^\prime}$ to be in the local subalgebra ${\cal H}$. The analogous  statement holds for the self-dual character of the generator $S_{-a_1a_2a_3}$  which  belong to the local subalgebra ${\cal H}$.  We observe  that half of the   equations   
${\cal E}_\alpha {}^{a^\prime } =0 $ and ${\cal E} _{[\alpha e_1e_2] }=0$ 
are equations of motion and half are  inverse Higgs conditions. 
\par
We now present a proposal for the {\bf non-linear } theory. We adopt equations (4.2.7) in the non-linear theory. Following the arguments in section seven  we conclude that  equation (4.2.7)  can be written as 
$$
\sqrt {-\gamma}\gamma^{\alpha \beta} \nabla_\beta  x^{\underline a} = -\epsilon^{\alpha \beta_1 \ldots \beta_5} \nabla _ {\beta_1 } x^{\underline a \underline b_1 \ldots \underline b_5 }\nabla _ {\beta_2 } x_{\underline b _2} \ldots \nabla _ {\beta_5 } x_{\underline b _5} 
\eqno(4.2.13)$$
It is tempting to write equation (4.2.8) in the non-linear theory as 
$ \nabla _{[a_1  }x_{  a_2a_3 ]} =0$ where $\nabla _{e  }= (s^{-1})_e{}^\alpha \nabla_\alpha$. However this equation is not invariant under the local subalgebra transformations due to the transformation of  $\delta ((s^{-1})_e{}^\alpha )\nabla_\alpha$. To address this matter we will consider that the brane,  which moves in the spacetime with coordinates $x^{\underline a}, \ x_{\underline a_1\underline a_2} , \ldots $, has an enlarged world volume which  is parameterised by $\xi^{\underline \alpha}=\{ \xi^\alpha , \xi^{\alpha_1 \alpha_2}\}$. This is natural given that it is moving in a much large spacetime that the usual eleven dimensional spacetime. We now consider the object 
$$
E_{\underline \alpha}{}^ A= \left(\matrix { 
\nabla_\alpha x^a & \nabla_\alpha x^{a_1a_2} \cr 
\nabla_{\alpha_1 \alpha _2 } x^a & \nabla_{\alpha_1 \alpha _2 }  x^{a_1a_2} \cr }\right)
\eqno(4.2.14)$$ 
In the non-linear theory we than replace equation (4.2.8) by 
$$
{\cal E}_{ a_1 a_2a_3}\equiv \nabla_{[ a_1 }x_{a_2a_3]} + {2\over 3} \nabla^{cd} x_{c[a_1| }\nabla_{d | a_1} x_{a_2a_3 ]}=0 
\eqno(4.2.15)$$
We note, using equation (4.2.4),   that under the $L_{ab^\prime}$ and $S_{-a_1a_2a_3}$ transformations of the local subalgebra 
$$
\delta(\nabla x^a)= -6 \Lambda_{+ }{}^{e_1e_2 a} \nabla x_{e_1e_2}=
-6 \Lambda_{+ }{}^{e_1e_2 a} E_{e_1e_2}
\eqno(4.2.16)$$ 
where we have used the conditions of equation (4.2.7). As a result we find that   
$$
\delta (\nabla _ {a_1a_2})= -6 \Lambda _{+ a_1a_2 }{}^{c}\nabla_c 
\eqno(4.2.17)$$ 
where  $\nabla _ {a_1a_2}=\equiv (E^{-1})_{a_1a_2}{}^{\underline \alpha} \nabla_{\underline \alpha}$.
Hence we find that 
$$
\delta ({\cal E}_{ a_1 a_2a_3}) = 
-2 {\cal E}_{ [a_1 | e_1 e_2} \Lambda _{+}{}^{e_1e_2 d} \nabla_d x_ {| a_1a_2]}+\ldots =0
\eqno(4.2.18)$$
which   is invariant. We have use the condition of equation (4.2.15)  and   we have discarded the terms in the variation that contain derivatives with respect to the higher level  world volume coordinates that parameterise the brane. The  $+\ldots$ indicates the presence of such terms. 
\par
The procedure we have used is similar to that used when constructing the field theory, that is, supergravity extended theories,  in E theory. The problem can be traced to the fact that the variation of $\nabla_{[ a_1 } x_{a_2a_3]}$ is not gauge invariant,  however, this is addressed in equation (4.2.15) by the addition of the second term involving a derivative with respect to the higher level coordinate $\xi_{a_1a_2}$ Thus we finds the same connection, as in the field theory case, between gauge symmetry and the presence of the  higher level coordinates. In the field theory one finds that such terms added are vital for the consistency of the theory even though the physical meaning of the higher level coordinates is still unclear. We present the above  non-linear theory only as a proposal as it would be good to examine its consistency to the same extent as has been done in the field theory case before being sure that  it is correct. 
\par
The part of  equation (4.2.15) that is an equation of motion, rather than an inverse Higgs condition, and does not involve derivatives with respect to the higher level brane coordinates  can be written in the non-linear theory as  
$$
\nabla _{[a_1  }x_{  a_2a_3 ]} + {1\over 3 !} \epsilon_{a_1 a_2 a_3}{}^{b_1b_2b_3}\nabla _{[b_1  }x_{  b_2b_3 ]}  =0
\eqno(4.2.19)$$
While this appears to be quite a simple equation its content is only apparent once one solves the inverse Higgs condition which obeys a   similar  condition but with the opposite duality. It would be interesting to see if the resulting equation which involves $x^{\underline a}$ and the gauge field $x_a^j$ agrees with the known dynamics for the M5 brane [35]. 
\medskip
{\bf 5. Branes in the  IIB  theory }
\medskip
In this section we will derive the equations of motion of the F1 and D1,  strings as well as those for the  D3 brane   that occur in the IIB theory. The IIB theory results from deleting node nine in the $E_{11}$ Dynkin diagram  and decomposing the $E_{11}\otimes_s l_1$ algebra in terms of the subalgebra of the remaining nodes, that is, $SL(10)\otimes SL(2)$ [34]. The IIB algebra in this decomposition can be found in reference [27]. We  will compute the dynamics in the absence of the background IIB supergravity fields and so we  will consider  the non-linear realisation of the semi-direct product of the Cartan involution subalgebra $I_c(E_{11})$ with the vector representation, that is,   $I_c(E_{11})\otimes_s l_1$. This algebra was worked out in collaboration with Michaella Pettit [33]. At level zero  $I_c(E_{11})$ is $SO(1,9)\otimes SO(2)$.
The generators of $I_c(E_{11})$ in this decomposition are 
$$
J^{\underline a}{}_{\underline b} , \  S ,  \   S^{{\underline a}_1{\underline a}_2}_{i} ,\ 
 S^{{\underline a}_1\ldots {\underline a}_4} , \  S^{{\underline a}_1\ldots {\underline a}_6}_{i} ,\ 
 S^{{\underline a}_1\ldots {\underline a}_8}_{i_1i_2} ,  \ S^{{\underline a}_1\ldots {\underline a}_7,\underline b}  ,\ldots 
\eqno{(5.1)}$$
where $i=1,2$ and $\underline a=0,1,\ldots 9$. Their definitions in terms of the underlying $E_{11}$ generators are given in equation (A.2) and 
$S^{{\underline a}_1\ldots {\underline a}_8}_{i_1i_2}= S^{{\underline a}_1\ldots {\underline a}_8}_{(i_1i_2)}$. The generators of the $l_1$ representations in this decomposition are given by 
$$
P_{\underline a} ; \ \ Z^{\underline a}_{i} ; \ \ Z^{{\underline a}_1{\underline a}_2{\underline a}_3} ; \ \ Z^{{\underline a}_1\ldots {\underline a}_5}_{i} ; \ \ Z^{{\underline a}_1\ldots {\underline a}_7}_{ij}; \ \ Z^{{\underline a}_1\ldots {\underline a}_7}, \ \ Z^{{\underline a}_1\ldots {\underline a}_6,\underline b} ,\ldots . 
\eqno(5.2)$$
  We raise and lower the $i,j,\ldots $ indices with $\delta _{i,j}$ but the indices $\underline a , \underline b , \ldots $ with the Minkowski metric corresponding to the fact that we are working with the algebra $SO(1,9)\otimes SO(2)$ at lowest level. 
\par
The group element used to construct  the non-linear realisation is of the form 
$g=g_l g_h$ 
where 
$$ 
g_l  = e^{x^{\underline a} P_{\underline a} + x_{\underline a}^{i}Z^{\underline a}_{i}  + x_{{\underline a}_1{\underline a}_2{\underline a}_3}Z^{{\underline a}_1{\underline a}_2{\underline a}_3} 
+ x_{{\underline a}_1\ldots {\underline a}_5}^{i} Z^{{\underline a}_1\ldots {\underline a}_5}_{i} + 
x_{{\underline a}_1\ldots {\underline a}_7}^{ij}Z^{{\underline a}_1\ldots {\underline a}_7}_{ i j}+ 
x_{{\underline a}_1\ldots {\underline a}_7} Z^{{\underline a}_1\ldots {\underline a}_7 } + x_{{\underline a}_1\ldots {\underline a}_6,\underline b}  Z^{{\underline a}_1\ldots {\underline a}_6,\underline b} +\ldots } 
\eqno(5.3)$$
while  $g_h$ belongs to the local subalgebra which depends on the brane being studied. 
\par
The Cartan forms can be written as  
$$
{\cal V}_l = g^{-1}_h ( dx^A  l_A )g_h=  
\nabla  x^{\underline a} P_{\underline a} + \nabla  x_{\underline a}^{i}Z^{\underline a}_{i}  + \nabla  x_{{\underline a}_1{\underline a}_2{\underline a}_3}Z^{{\underline a}_1{\underline a}_2{\underline a}_3} 
$$
$$
+ \nabla x_{ {\underline a}_1\ldots {\underline a}_5}^{i} Z^{{\underline a}_1\ldots {\underline a}_5}_{i}  +
\nabla  x_{{\underline a}_1\ldots {\underline a}_7}^{ ij}Z^{{\underline a}_1\ldots {\underline a}_7}_{ ij}+ 
\nabla _\alpha x_{{\underline a}_1\ldots {\underline a}_7} Z^{{\underline a}_1\ldots {\underline a}_7 } + \nabla  x_{{\underline a}_1\ldots {\underline a}_6,\underline b}Z^{{\underline a}_1\ldots {\underline a}_6,\underline b}+\ldots 
\eqno(5.4)$$
where $\nabla = d\xi^{\underline \alpha} \nabla _{\underline \alpha} $\par
Using equation (2.15) the transformations of the Cartan forms under a local transformation which involves the most general possible $h\in  I_c(E_{11})$, that is,  \quad
$h= 1-(-J_{\underline a_1\underline a_2}\Lambda ^{\underline a_1\underline a_2}+S^{\underline a_1\underline a_2}_{i} \Lambda _{\underline a_1\underline a_2}^{i} 
+S^{\underline a_1\ldots \underline a_4}\Lambda _{\underline a_1\ldots \underline a_4})$ are given  found to be 
$$
\delta {\cal V}_l= [ -J_{\underline a_1\underline a_2}\Lambda ^{\underline a_1\underline a_2}+S^{\underline a_1\underline a_2}_{i} \Lambda _{\underline a_1\underline a_2}^{i} 
+S^{\underline a_1\ldots \underline a_4}\Lambda _{\underline a_1\ldots \underline a_4} + \Lambda S, {\cal V}_l]  
\eqno(5.5)$$
which results, using reference [33] in the variations 
$$
\delta(\nabla_\alpha x^{\underline a})= -4\Lambda^{\underline a \underline b}_i \nabla_\alpha x_{\underline b}^i + 2\nabla _\alpha x^{\underline b} \Lambda _{\underline b}{}^{\underline a} +48 \nabla _\alpha x^{\underline b_1 \underline b_2 \underline b_3}\Lambda_{\underline b_1 \underline b_2 \underline b_3}{}^{\underline a} \  
$$
$$
\delta(\nabla_\alpha x_{\underline a}^i)= -\Lambda^{\underline a \underline b}_i \nabla _\alpha x^{\underline b}+ 2 \nabla _\alpha x_{\underline b}^i \Lambda ^{\underline b}{}_{\underline a}-6 \nabla _\alpha x^{\underline b_1 \underline b_2 \underline a}\epsilon _{ij} \Lambda_{\underline b_1 \underline b_2}^j +120 \nabla _\alpha x^i_{\underline b_1\ldots\underline b_4\underline a} \Lambda ^{\underline b_1\ldots\underline b_4} 
-{1\over 2} \nabla_\alpha x^j_{\underline a}\epsilon_{j}{}^{i}\Lambda
$$
$$
\delta(\nabla_\alpha x_{\underline a_1 \underline a_2\underline a_3})= 
6\nabla_\alpha x_{\underline b [\underline a_2\underline a_3}\Lambda ^{\underline b}{}_{\underline a_1]}
+2 \nabla_\alpha x^{\underline b} \Lambda _{\underline b \underline a_1\underline a_2\underline a_3}
-20 \nabla _\alpha x^i_{\underline b_1 \underline b_2 \underline a_1\underline a_2\underline a_3}\Lambda _i^{\underline b_1 \underline b_2 } -\nabla _\alpha x_{[\underline a_1}^j\epsilon_{ij} \Lambda ^i_{\underline a_2\underline a_3 ]}
$$
$$
\delta(\nabla_\alpha x_{\underline a_1 \ldots \underline a_5}^i)= 
10\nabla_\alpha x_{\underline b [\underline a_2\ldots \underline a_5}^i\Lambda ^{\underline b}{}_{\underline a_1]}-\Lambda _{[\underline a_1\ldots \underline a_4}\nabla_\alpha x^i_{\underline a_5]} 
+\Lambda^i_{[ \underline a_1\underline a_2}\nabla_\alpha x_{\underline a_3\underline a_4\underline a_5]} -{1\over 2} \nabla_\alpha x^j_{\underline a_1\ldots \underline a_5}\epsilon_{j}{}^{i}\Lambda
+\ldots 
\eqno(5.6)$$
However, it is important to remember that the local subgroup ${\cal H}$ is a subgroup of  $I_c(E_{11})$ and so the above parameters must be suitably  restricted so that the group element $h$ belongs to the chosen local subgroup ${\cal H}$ for the brane we are considering.  
\medskip
{\bf 5.1 The IIB string  }
\medskip
In this subsection we will derive  the equation of motion of the F1 and D1 strings in the IIB theory. The corresponding charges,  $Z^{\underline a}_i$
are an SL(2) doublet. We can treat the dynamics for both strings simultaneously by introducing two constants $q^i, i=1,2$ which are normalised so as to obey 
$q^2=q^iq^i=1$. We take the charge of the string we are considering to be given by $< Z^{\underline a}_i >=q_i Z^a$. Looking at the group element $g_l$ of equation (5.3) we find that the dynamics of string we are considering will contain the level zero coordinate $x^{\underline a}$, corresponding to $P_{\underline a}$   and  coordinate  $y_{\underline a}\equiv 2 q_i  x^i_{\underline a}$ where we have introduced a factor of 2 for reasons that will become apparent. To define the orthogonal compliment to $q^i$ we introduce 
$\bar q^i= \epsilon^{ij} q_j$. We note that $\bar q^iq_i=0$ and 
$\bar q^2=\bar q ^i\bar q^i=1$. The orthogonal  level one coordinate can be taken to be $z_{\underline a}\equiv 2\bar q_i  x^i_{\underline a}$ and we can write $x^i _{\underline a} ={1\over 2} q^i y_ {\underline a} +{1\over 2} \bar q^i z_ {\underline a}$.  
\par
We must now choose the local subgroup ${\cal H}$ that will lead to the string dynamics. Rather than just write it down we will now motive our choice. Examining equation (5.6) we find that if $S$ was in the local subalgebra ${\cal H}$ it would induce a transformation with parameter $\Lambda $ that interchanges the Cartan from for $y_{\underline a}$ with  that for $z_{\underline a}$. In terms of the charges it interchanges the charge $q^iZ_i^a$ with $\bar q^iZ_i^a$. As we want a dynamics that only contains the coordinates, or charges,  associated with the chosen string   we do not take the generators $S$ to be in the local subgroup ${\cal H}$. 
\par
Examining the first of the equations in (5.6) we realise that under a  transformation with parameter $\Lambda^{\underline a \underline b}_i$ the above Cartan forms  transform as 
$$\delta (\nabla _\alpha x^{\underline a} )= -2 (q^i\Lambda^{\underline a \underline b}_i \nabla _\alpha y_{\underline b} + \bar q^i \Lambda^{\underline a \underline b}_i \nabla _\alpha z_{\underline b})$$
and  
$$\delta (\nabla _\alpha y^{\underline a} )= -{2}q^i \Lambda^{\underline a \underline b}_i\nabla _\alpha x_{\underline b}\ ,\ \  \delta (\nabla _\alpha z^{\underline a} )= -{ 2}\bar q^i \Lambda^{\underline a \underline b}_i\nabla _\alpha x_{\underline b}$$
Since the chosen string should involve the Cartan forms for the coordinates $x^{\underline a}$ and $y^{\underline a}$  and not that for $z^{\underline a}$ we should take the parameter $\Lambda^{\underline a \underline b}_i$ to be in the form $\Lambda^{\underline a \underline b}_i= - q_i\tilde \Lambda^{\underline a \underline b}$ and so the generators 
$\hat S^{\underline a_1\underline a_2}\equiv q_i S^{\underline a_1 \underline a_2 i }$  to be in ${\cal H}$ and  we exclude the generator $\bar q^i S^{\underline a_1 \underline a_2}_i$ from ${\cal H}$. Using similar arguments one concludes that the generator $S^{{\underline a}_1\ldots {\underline a}_4} $ is not in ${\cal H}$
\par 
As a result we consider the local subgroup ${\cal H}$ should contain generators that are taken from the set  
$$
\{ J_{\underline a\underline b},\     \hat   S^{\underline a_1\underline a_2}  , \ldots \}
\eqno(5.7)$$
We note that at this point we have not chosen the local subalgebra ${\cal H}$ only stated that it is to be a subgroup of the generators listed in equation (5.7). 
Using the results of appendix B we find that   these generators obey the  commutators  
$$
[ \hat S^{\underline a_1\underline a_2} , \hat S^{\underline b_1\underline b_2} ]= -4 \delta^{[\underline a_1 } _{ [\underline b_1}J^{\underline a_2]}{}_{\underline b_2]} , \ 
[  J^{\underline a_1\underline a_2} , \hat S^{\underline b_1\underline b_2} ]= -4 \delta^{[\underline a_1 } _{ [\underline b_1}\hat S^{\underline a_2]}{}_{\underline b_2]} , \ 
[  J^{\underline a_1\underline a_2} , J^{\underline b_1\underline b_2} ]= -4 \delta^{[\underline a_1 } _{ [\underline b_1} J^{\underline a_2]}{}_{\underline b_2]}  
\eqno(5.8)$$
We recognise this subalgebra as the $O(10)\otimes O(10)$ as given in equation (3.6). 
\par
The transformations of the Cartan forms of equation (5.5) corresponding to  generators $\hat S^{\underline a_1\underline a_2}$ and the Lorentz transformations are given by equation (5.6) with parameter $\tilde \Lambda _{\underline a_1\underline a_2} $ are given by 
 $$
\delta(\nabla_\alpha x^{\underline a})= 2\nabla_\alpha  y_{\underline b}\tilde \Lambda^{\underline b \underline a }  + 2\nabla _\alpha x^{\underline b} \Lambda _{\underline b}{}^{\underline a} ,
\eqno(5.9)$$
$$
\delta(\nabla_\alpha y^{\underline a})= 2\nabla_\alpha  y_{\underline b}\tilde \Lambda^{\underline b \underline a }  + 2\nabla _\alpha y^{\underline b} \Lambda _{\underline b}{}^{\underline a} ,
\eqno(5.10)$$
as well as  
$$
\delta(\nabla_\alpha z_{\underline a})=  2 \nabla _\alpha z_{\underline b} \Lambda ^{\underline b}{}_{\underline a}+6 \nabla _\alpha x^{\underline b_1 \underline b_2 \underline a}\tilde  \Lambda_{\underline b_1 \underline b_2}, \ldots  
\eqno(5.11)$$
$$
\delta(\nabla_\alpha x_{\underline a_1 \underline a_2\underline a_3})= 
6\nabla_\alpha x_{\underline b [\underline a_2\underline a_3}\Lambda ^{\underline b}{}_{\underline a_1]}
+\nabla _\alpha z_{[\underline a_1}\tilde  \Lambda _{\underline a_2\underline a_3 ]}+\ldots 
\eqno(5.12)$$
where $+\ldots $ denote higher level coordinates
\par
Clearly,  we may  set $\nabla_\alpha z_{\underline a}=0$ and $\nabla_\alpha x_{\underline a_1 \underline a_2\underline a_3}=0$ while preserving the symmetries of the non-linear realisation. This is the choice we now adopt. 
\par
The reader will have realised that we have arrived at exactly the same situation as we had in section four where we studied the IIA string. Indeed the generators of equation (5.7) obey the same SO(10,10)  algebra as those of equation (3.5) and the coordinates $x^{\underline a}$ and $y^{\underline a}$ can be identified with the coordinates of the same name  in section four. The derivation of the dynamics then proceeds just as in section four with the local subalgebra ${\cal H}$ being that of equation (3.8) with $S_{ab}$ now being $\hat S_{ab}$ etc. The equations of motion are  equation (3.21), and finally  equation (3.26).  
\medskip
{\bf 5.2 The  D3  brane}
\medskip
The D3  brane  possess  a four dimensional world volume and so we divide the indices into their longitudinal and transverse parts, in particular    
$\underline a=0,1,\ldots ,9$  divides into $a=0,1, 2,3$ and $a^\prime = 4,5,\ldots ,9$. The generators of the vector representation are given in equation (5.2) and those of $I_c(E_{11})$ in equation (5.1). The algebra $I_c(E_{11})\otimes_s l_1 $ can be found in appendix B. The non-linear realisation of $I_c(E_{11})\otimes_s l_1$ 
 is constructed from the group element $g=g_l g_h$ where $g_l$ can be found in equation (5.3). 
\par
We choose the local subgroup ${\cal H}$ to be given by 
$$
{\cal H}= \{J_{ab}, J_{a^\prime b^\prime} ,  S, L_{ab^\prime} , 
  L^{i+}_{a_1a_2} ,\hat S  , \ldots \}
\eqno(5.13)$$
where 
$$
\hat S\equiv \epsilon^{a_1\ldots a_4} S_{a_1\ldots a_4} ,\quad L_{ab^\prime} \equiv 3!J_{ab^\prime}+ \epsilon _{a}{}^{e_1e_2e_3} S_ {e_1e_2e_3 b^\prime} , \ 
 L^{i\pm}_{a_1a_2}\equiv S^{i}_{a_1a_2}\pm{1\over 2} \epsilon^{ij} \epsilon_{a_1a_2}{}^{b_1b_2}  S_{ b_1b_2 j}
\eqno(5.14)$$
We note that ${\cal H}$ contains the SO(2) generator $S$ which is part of the SL(2) symmetry of IIB theory. 
\par
We now give the algebra that the generators of ${\cal H}$ obey beginning with the commutators  that involve the generators $S$  which are given by 
$$ 
  [{S}, J_{ab}] =0 , \ \  [{S}, J_{a b'}] =0 , \ \  [{S}, L_{cd'}] =0 , \ \ [S, \hat S] =0 , \ \ [S, L^{i+}_{a_1a_2} ] =-{1\over 2} \epsilon_{ij} L^{j+} _{a_1a_2}, \ldots 
\eqno(5.15)$$
while, those that involve the generators  $\hat S$ are
$$
[{\hat S}, J_{ab}] =0  ,\ \ 
[{\hat S}, J_{a^\prime b^\prime}] =0  , \ \ [{\hat S}, L_{cd'}] = -4!L_{cd'}  , \ \ [\hat S, L^{i+}_{a_1a_2} ] = 24  L^{i+}_{a_1a_2} ,\ldots  
\eqno(5.16)$$
The commutators involving the generators $L_{ab^\prime}$ are given by 
$$ 
[L_{ab'}, L_{cd'}] = 0 ,  \ \ [L_{a_1a_2}^{i+}, L_{cd'}] =  0  , \ \ 
, \ldots 
\eqno(5.17)$$
The commutators involving the generators $L_{a_1a_2}^{i+}$ are given by 
$$
 [L^{i+}_{a_1a_2}, L^{j+}_{b_1b_2}] = 0, \ldots 
\eqno(5.18) $$
\par
As noted in section two, equation (2.12) we  can use the local symmetry to choose the parts of $g_h$ that are in ${\cal H}$ to vanish. Indeed we can choose  the  generators which are in $I_c(E_{11})$ but not in ${\cal H}$, up to level two,  to be  $J^a{}_{b^\prime}$,  $L^{i-}_{a_1a_2}$, $S^i_{ab^\prime}$ and  $S^i_{a^\prime b^\prime}$ and as a result we may choose 
$g_h$ to be of the form 
$$
g_h= e^{-\varphi _a{}^{b^\prime} J^a{}_{b^\prime}}e^{\varphi^{i+}{}^{a_1a_2}
L^{i-}_{a_1a_2}}e^{S^i_{ab^\prime} \varphi^{ab^\prime}_i}e^{S^i_{a^\prime b^\prime} \varphi^{a^\prime b^\prime}_i}\ldots 
\eqno(5.19)$$
where 
$$
\varphi_{\pm}{}_{a_1a_2 i}\equiv {1\over 2}(\varphi^{i}_{a_1a_2}
\pm{1\over 2} \epsilon^{ij} \epsilon_{a_1a_2}{}^{b_1b_2}  \varphi_{ b_1b_2 j} )
= \pm {1\over 2} \epsilon^{ij} \epsilon_{a_1a_2}{}^{b_1b_2}  \varphi_{ b_1b_2 j} 
\eqno(5.20)$$
We note that $T^{i+}{}^{a_1a_2} R^{i-}_{a_1a_2}=0$ for any two objects 
$T^{i+}{}^{a_1a_2}$ and $R^{i-}_{a_1a_2}$ where the $\pm$ projections are as expected. 
\par
We now choose an alternative basis of the vector representation to that given in equation (5.2) that is suited to the action of the local subgroup ${\cal H}$. We take the basis 
$$
N_a^{\pm} , P_{a^\prime} , Z_a^i , N_{a^\prime} ^{i \pm} , Z^{a_1a_2b^\prime} , Z^{ab_1^\prime b_2^\prime} , Z^{a_1^\prime a_2^\prime a_3^\prime} , \ldots 
\eqno(5.21)$$
where 
$$
N_a^{\pm} = P_a\pm {1\over 2.3!} \epsilon _{a e_1e_2e_3}Z^{e_1e_2e_3} ,
 N_{a^\prime} ^{i \pm}=  Z_{a^\prime} ^{i }\mp {1\over 4!}\epsilon^{e_1\ldots e_4 } Z^i_{e_1\ldots e_4 a^\prime } 
\eqno(5.22)$$
\par
We now present the  commutators of the vector representation with the generators of ${\cal H}$. The action of the Lorentz generators is standard and so we begin with the commutators  containing  $S$:

$$ [S, P_{a'}] = 0 , \ \ [S, N^{\pm}_a] = 0, \ \ [S, Z^{a_1a_2b'}] = 0, \ \ [S, Z^{ab'_1b'_2}] = 0, $$
$$ [S, N^{\pm i}_{a'}] = -{1\over2}\varepsilon^{ij}N^{\pm j}_{a'} , \ \ [S, Z_c^i] = -{1\over2}\varepsilon^{ij}Z^j_c 
\eqno(5.23)$$
The commutators with the generators  $L_{ab'}$ are given by 
$$
[L_{ab'}, N_c^-] = 0, \ \  [L_{ab'}, N_c^+] = -2.3!\eta_{ac}P_{b'} + 3\varepsilon_{ace_1e_2}Z^{e_1e_2b'} 
$$
$$
[L_{ab'}, P_{c'}] = 3!\eta_{b'c'}N^-_a, \ \
[L_{ab'}, Z^i_c ]=-3! \eta_{ac} N^-_i{}^{b^\prime}
$$

$$
[L_{ab'}, N^{-}_{ic'}] = 0, [L_{ab'}, N^{+}_{ic'}] = 2.3!\eta_{b'c'}Z_a^i - 2\varepsilon_{ae_1e_2e_3}Z^{e_1e_2e_3b'c'}_i , $$

$$ [L_{ab'}, Z_{c_1c_2c_3}] = -3.3!\eta_{a[c_1|}Z_{b'|c_2c_3]}+12\varepsilon_{ac_1c_2c_3}P_{b'} , 
$$
$$ [L_{ab'}, Z_{c_1c_2d'}] = 12\eta_{a[c_1|}Z_{|c_2] b'd'} - 12\eta_{b'd'}\varepsilon_{ac_1c_2e}N_e^- , $$
$$ [L_{ab'}, Z_{c'_1c'_2c'_3}] = 18\eta_{b'[c'_1}Z_{c'_2c'_3]a}, \ \  [L_{ab'}, Z^{cd'_1d'_2}] =- 3!\eta_{ac}Z_{b'd'_1d'_2} -12\eta_{b'[d'_1}Z_{d'_2]ac} 
\eqno(5.24)$$

\par
The commutators of $L^{i+}_{a_1a_2}$ with the vector representation  generators are given by $$
[L_{a_1a_2}^{i+}, N_b^-] = 0, \ \ [L_{a_1a_2}^{i+}, N^+_b] = \varepsilon^{ij}\varepsilon_{ba_1a_2c}Z^j_c + 2\eta_{b[a_1}Z^i_{a_2]} $$
$$[L_{a_1a_2}^{i+}, P_{b'}] = 0, \ \ [L_{a_1a_2}^{i+}, Z_b^j] = 2\varepsilon^{ij}\varepsilon^{a_1a_2bd}N_d^- + 4\delta^{ij}\eta_{b[a_1}N^-_{a_2]} , $$
$$[L_{a_1a_2}^{i+}, Z_{b'}^j] = -\varepsilon^{ij}Z^{a_1a_2b'} - {1\over2}\delta^{i,j}\varepsilon_{a_1a_2c_1c_2}Z^{c_1c_2b'},
$$
$$ [L^{i+}_{a_1a_2}, N^{-j}_{b'}] = 0, \ \ [L_{a_1a_2}^{i+}, N_{b'}^{+j}] = -\delta^{i,j}\varepsilon_{a_1a_2c_1c_2}Z^{c_1c_2b'} - 2\varepsilon^{ij}Z^{a_1a_2b'} 
$$ 

$$ \ \ [L_{a_1a_2}^{i+}, Z^{b_1b_2b_3}] = 6\varepsilon^{ij}\delta_{a_1a_2}^{[b_1b_2}Z^{b_3]}_j - 3\delta^i_j\varepsilon_{a_1a_2[b_1b_2|}Z_{j|b_3]} 
$$
$$[L_{a_1a_2}^{i+}, Z^{b_1b_2c'}] = -\varepsilon_{a_1a_2b_1b_2}N^{i}{}_{c'}^- + 2\varepsilon^{ij}\delta^{b_1b_2}_{a_1a_2}N_j^{-c'} ,
$$
$$ \ \ [L_{a_1a_2}^{i+}, 
N_-^{cd'_1d'_2}] =  2Z^i{}^{a_1a_2cd'_1d'_2} +\varepsilon^{ij}\varepsilon_{a_1a_2b_1b_2}Z^{b_1b_2cd'_1d'_2}_j ,
\ \ [L_{a_1a_2}^{i+}, 
N_+^{cd'_1d'_2}] = 0$$
$$ 
[L_{a_1a_2}^{i+}, Z^{c'_1c'_2c'_3}] =  Z^i{}^{a_1a_2c'_1c'_2c'_3} + {1\over2}\varepsilon^{ij}\varepsilon_{a_1a_2b_1b_2}Z^{b_1b_2c'_1c'_2c'_3}_j, 
\eqno(5.25)$$
\par
Taking the commutators with $\hat{S} = \varepsilon^{a_1\ldots a_4}S_{a_1\ldots a_4}$, we find the following 
$$ [\hat{S}, P_{a'}] = 0 , \ \ [\hat{S}, N_a^{\pm}] = \pm4.3!N^{\pm}_b, \ \ [\hat{S}, N^{\pm i}_{b'}] = \pm 4!N_i^{\pm b'} , 
 \ \ [\hat{S}, Z^i_c] = 0 ,
$$ 
$$ [{\hat S}, N_\pm ^{c_1c_2d'}] = \pm N^{cd'_1d'_2}_{\pm} 
 , \ \  [ \hat{S}, Z^{b_1b_2c'}] = 0 , \ \ [ \hat{S}, Z^{bc'_1c'_2}] = 0 
, \ \ [ \hat{S}, Z^{c'_1c'_2 c'_3}] = 0
\eqno(5.26)$$
Examining the above commutators we find that 
$$
N_c^-, \ N_{ic^\prime} ^- , \ldots $$
is an irreducible representation under the local subalgebra ${\cal H}$. 
\par
When written in terms of the basis of equation (5.21) the Cartan forms are given by 
$$ {\cal V} = {\cal E}^{\pm}{}^a N_a^{\pm} + {\cal E}^{c'}P_{c'} + N^{\pm i}_{c'}{\cal E}^{\pm c'}{}_i + {\cal E}_i{}^c Z^i_c+{\cal E}_{a_1a_2b'}Z^{a_1a_2b'}  + {\cal E}_{ab'_1b'_2} Z^{ab'_1b'_2} + {\cal E}_{a'_1a'_2a'_3}Z^{a'_1a'_2a'_3} +\ldots 
\eqno(5.27)$$
where the $\pm$ are summed over and 
$$
{\cal E}^{\pm}{}^a = {1\over2} \nabla x^a \mp \varepsilon^{ab_1b_2b_3}\nabla x_{b_1b_2b_3}, \   {\cal E}^{a'} = \nabla x^{a'} , \  {\cal E}_a{}^i = \nabla x_a{}^i ,\ 
$$
$$ 
{\cal E} ^{\pm a'}{}^{i} = {1\over 2}(\nabla x_i^{a'} \pm 5\varepsilon_{e_1\ldots e_4} x^{e_1\ldots e_4 a'}{}_i) ,\  
$$
$$
{\cal E} ^{a_1a_2b^\prime} = \nabla x^{a_1a_2b^\prime},\ \ 
{\cal E}^{cd_1^\prime d_2^\prime }=  \nabla x^{cd_1^\prime d_2^\prime } , \ \ {\cal E}^{d_1^\prime d_2^\prime d_3^\prime }=  \nabla x^{d_1^\prime d_2^\prime d_3^\prime }
\eqno(5.28)$$ 
In these equations the ${\cal E}$'s are forms, that is, ${\cal E}^\bullet = d\xi^{\underline \alpha }{\cal E}_{\underline \alpha}^\bullet $ and similarly $\nabla =d\xi^{\underline \alpha} {\nabla }_{\underline \alpha} $. The variation of the Cartan forms under the local ${\cal H}$ transformations is given by 
$$
 \delta {\cal V} = [\hat{\Lambda}\hat{S} + \Lambda S + \Lambda^{ab'}L_{ab'} + \Lambda^{-a_1 a_2 i}L_{+}{}_{a_1a_2 i} , {\cal V} ] 
\eqno(5.29)$$
and we find that 
$$  \delta {\cal E}^+{}_a = 4! {\cal E} ^+{}_a\hat \Lambda ,\ \ 
\delta {\cal E}_{b'}= - 12{\cal E}^+{}_a\Lambda^{a}{}_{b'} ,\ \ 
\eqno(5.30)$$
$$
 \delta {\cal E}_a{}^i = {\Lambda \over 2}\varepsilon^{ij}{\cal E}^a {}_j+ 12 \Lambda^{ab'}{\cal E}^{+}{}_{b'}{}^{i}  + 4{\cal E}_c^+\Lambda_{-}{}^{c}{}_{ai} 
,\ \  \delta {\cal E}^{+}{}^{i}{}_{a'} = {\Lambda \over 2}{\cal E}^+{}_{a'j}\varepsilon^{ij} + 4{\hat \Lambda}{\cal E}^{+}{}_{a'}{}^{i}  
\eqno(5.31)$$
$$ 
\delta {\cal E}_{a_1a_2b'} = 3\Lambda^{db'}\varepsilon_{dca_1a_2}{\cal E}^+{}_c + 12{\cal E}{}_{[ a_1| d'b'}\Lambda_{| a_2 ]}{}^{d'}  - 4\varepsilon^{ij}\Lambda_{- a_1a_2}{}_{i}{\cal E}^{+}{}_{b'j} 
$$
$$ \delta {\cal E}_{a'_1a'_2a'_3} = -3!\Lambda_{e[a'_1|}{\cal E}^{\pm}{}_{e|a'_2a'_3]} , \ \  \delta {\cal E}^{ab'_1b'_2} = -12\varepsilon^{da[b'_1|}\Lambda^{d|b'_2]} + 18 {\cal E}^{d^\prime b_1^\prime b_2^\prime} \Lambda ^{a}{}_{d^\prime}  
\eqno(5.32)$$
$$ \delta {\cal E}^- {}_a= -4!{\cal E}^-{}_a{\hat \Lambda} + 3!{\cal E}_{b'}\Lambda^{ab'} - 12\varepsilon_{dc_1c_2a}\Lambda^{ab'}{\cal E}^{c_1c_2b'} + 8{\cal E}^{+bi}\Lambda_{-ba i }
$$
$$ \delta {\cal E}^{-i}{}_{a'} = -3!{\cal E}_d{}^i\Lambda^{da'} - 4{\hat \Lambda}{\cal E}^{-}{}_{a'}{}^{i} + {\Lambda \over 2}{\cal E}^-{}_{a'j}\varepsilon^{ij}- 4\varepsilon^{ij}\Lambda_{-c_1c_2}{}^j{\cal E}^{c_1c_2a'} 
\eqno(5.33)$$

\par
Examining equation (5.30) we find that we can consistently set to zero 
$$ {\cal E}_\alpha {}^{a'}= \nabla_\alpha x^{a'} = 0 ,\ \ {\cal E}_\alpha {}^{+ a }= {1\over2} \nabla_\alpha x^a - \varepsilon^{ab_1b_2b_3}\nabla _\alpha x^{b_1b_2b_3}= 0
\eqno(5.34)$$
that is, we can set these constraints and those of equation (5.33) to zero while preserving all the symmetries of the non-linear realisation. 
\par
We now restrict the discussion to the {\bf linearised } theory.   
Examining the above local  variations we see that we can set 
$$
{\cal E}^L{}_{[\alpha }{}_{a ]}{}^{i} =0,\ \ 
\eqno(5.35)$$
as well as  
$$
{\cal E}_\alpha {} ^{+ a^\prime i}=0
\eqno(5.36)$$
and preserve all the symmetries of the non-linear realisation. 
\par
To find the meaning  of the above conditions for the linearised theory we compute their expressions in terms of the fields. As a first step we calculate the Cartan forms for a general group element,  $g_h$, that is, one that belongs to $I_c(E_{11})$ and so has the form  
$$
g_h= e^{ -J_{\underline a_1\underline a_2}\phi ^{\underline a_1\underline a_2}+S^{\underline a_1\underline a_2}_{i} \phi _{\underline a_1\underline a_2}^{i} 
+S^{\underline a_1\ldots \underline a_4}\phi _{\underline a_1\ldots \underline a_4} + \phi S}  
\eqno(5.37)$$
Proceeding in this way we have not yet used the local subalgebra ${\cal H}$ to restrict the group element $g_h$ and so the  fields $\phi$. 
The result at the linearised level is given by 
$$
\nabla_\alpha x^{\underline a}=\partial_\alpha x^{\underline a} -4\phi^{\underline a \underline b}_i \partial_\alpha x_{\underline b}^i + 2\partial _\alpha x^{\underline b} \phi _{\underline b}{}^{\underline a} +48 \partial _\alpha x^{\underline b_1 \underline b_2 \underline b_3}\phi_{\underline b_1 \underline b_2 \underline b_3}{}^{\underline a} \  +\ldots
\eqno(5.38)$$
$$
\nabla_\alpha x_{\underline a}^i =\partial_\alpha x_{\underline a}^i -\phi^{\underline a \underline b}_i \partial _\alpha x^{\underline b}+ 2 \partial _\alpha x_{\underline b}^i \phi ^{\underline b}{}_{\underline a}-6 \partial _\alpha x^{\underline b_1 \underline b_2 \underline a}\epsilon _{ij} \phi_{\underline b_1 \underline b_2}^j +120 \partial _\alpha x^i_{\underline b_1\ldots\underline b_4\underline a} \phi ^{\underline b_1\ldots\underline b_4} 
-{1\over 2} \partial_\alpha x^j_{\underline a}\epsilon_{j}{}^{i}\phi  +\ldots
\eqno(5.37)$$
$$
\nabla_\alpha x_{\underline a_1 \underline a_2\underline a_3}= 
\partial_\alpha x_{\underline a_1 \underline a_2\underline a_3}+
6\partial_\alpha x_{\underline b [\underline a_2\underline a_3}\phi ^{\underline b}{}_{\underline a_1]}
+2 \partial_\alpha x^{\underline b} \phi _{\underline b \underline a_1\underline a_2\underline a_3}
-20 \partial _\alpha x^i_{\underline b_1 \underline b_2 \underline a_1\underline a_2\underline a_3}\phi _i^{\underline b_1 \underline b_2 } -\partial _\alpha x_{[\underline a_1}^j\epsilon_{ij} \phi ^i_{\underline a_2\underline a_3 ]} +\ldots
\eqno(5.38)$$
$$
\nabla_\alpha x_{\underline a_1 \ldots \underline a_5}^i= 
\partial_\alpha x_{\underline a_1 \ldots \underline a_5}^i+
10\partial_\alpha x_{\underline b [\underline a_2\ldots \underline a_5}^i\phi ^{\underline b}{}_{\underline a_1]}-\phi _{[\underline a_1\ldots \underline a_4}\partial_\alpha x^i_{\underline a_5]} 
+\phi^i_{[ \underline a_1\underline a_2}\partial_\alpha x_{\underline a_3\underline a_4\underline a_5]} -{1\over 2} \partial_\alpha x^j_{\underline a_1\ldots \underline a_5}\epsilon_{j}{}^{i}\phi
+\ldots 
\eqno(5.39)$$
where $+\ldots$ means terms that are higher level in the fields $\phi$ and $x$. 
\par
If we choose static gauge $\partial _\alpha x^a= \delta_\alpha ^a$ then as zeroth order we conclude from equation (5.34) that 
$\partial_\alpha x_{a_1a_2a_3}= -{1\over 2.3!}\epsilon_{\alpha a_1a_2a_3}$. 
Whereupon the find that to at most either first order in the fields $x$ or  $\phi$ that 
$$
\nabla_\alpha x^{a^\prime }=\partial_\alpha x^{a^\prime } +2\phi_\alpha {}^{a^\prime}-4\epsilon _{\alpha e_1e_2e_3} \phi^{e_1e_2e_3 a^\prime}+\ldots 
\eqno(5.40)$$
$$
\nabla_\alpha x_{ a_1  a_2 b^\prime}= 
\partial_\alpha x_{ a_1  a_2 b^\prime}+2\phi_{\alpha a_1a_2 b^\prime} -{1\over 6} \epsilon _{\alpha a_1a_2 e} \phi^e{}_{b^\prime} 
+\ldots 
\eqno(5.41)$$
We observe that 
$$
\nabla_\alpha x^{a^\prime }+2 \epsilon_{ \alpha}{}^{ \beta e_1 e_2}\nabla_\beta x_{e_1e_2 }{}^{a^\prime} 
= \partial_\alpha x^{a^\prime }+2 \epsilon_{ \alpha}{}^{ \beta e_1 e_2}\partial_\beta x_{e_1e_2 }{}^{a^\prime}
\eqno(5.42)$$
while 
$$
\nabla_\alpha x^{a^\prime }-2 \epsilon_{ \alpha}{}^{ \beta e_1 e_2}\nabla_\beta x_{e_1e_2 }{}^{a^\prime} 
= \partial_\alpha x^{a^\prime }-2 \epsilon_{ \alpha}{}^{ \beta e_1 e_2}\partial_\beta x_{e_1e_2 }{}^{a^\prime} 
+2(\phi_\alpha {}^{a^\prime}-4\epsilon _{\alpha e_1e_2e_3} \phi^{e_1e_2e_3 a^\prime})
\eqno(5.43)$$
\par
The constraint ${\cal E}_\alpha {}^{a'}=0$  of equation (5.34) and ${\cal E}_\alpha {}^{+ a_1a_2  b^\prime }=0$ of equation (5.35) imply that the objects in equations (5.42) and (5.43) vanish. Implementing this we see that  equation (5.43) allows is to solve for the combination of $\partial x$' s that occurs in terms of the fields $\phi$'s and is an inverse Higgs condition while equation (5.43) is an equation of motion. Indeed taking a derivative it is the correct equation of motion for the D3 brane at the linearised level. 
\par
 As explained in the context of the other branes our choice of local subalgebra is so as to ensure that only the combination of $\phi$'s that occurs in equation (3.43) appears in the group element $g_h$ once we have used the local subalgebra to restrict the $\phi$'s in $g_h$. 
\par
We now investigate the Cartan form that contains the world volume vector  $x_a^i$ in the linearised theory, for terms at most linear in either $x$ and $\phi$'s it  takes the form 
$$
\nabla_{[ \alpha}  x_{ a] }{}^{ i }=\partial_{[\alpha }x_{ a ]}{}^{ i} 
+4\phi_+{}_{\alpha  a}{}^i
\eqno(5.44)$$
where $\phi_{\pm}{}_{\alpha  a}{}^i= {1\over 2}(\phi^{}_{\alpha  a}{}^i
\pm{1\over 2} \epsilon_{\alpha a e_1e_2} \epsilon^{ij}\phi^{e_1e_2 }{}^j )$. 
Using the constraint ${\cal E}_{[\alpha} {}_{a] }{}^{i} =0$ of equation (5.35)  we find that 
$$
\partial_{[\alpha } x_{ a ]}{}^{ i }-{1\over 2}\epsilon_{\alpha a}{}^{\beta e} \epsilon^{ij}
\partial_\beta x_{e}{}_j=0
\eqno(5.45)$$
and 
$$
\partial_{[ \alpha } x_{ a ]}{}^{ i }+{1\over 2} \epsilon_{\alpha a}{}^{\beta e} \epsilon^{ij}
\partial_\beta x_{e}{}_j+8\phi^+{}_{\alpha  a}{}^i=0
\eqno(5.45)$$
The last equations just express the fact that the anti-self-dual part of the field strength 
$f_{ab}{}^i= \partial_a x_b{}^i- \partial_b x_a{}^i$ is expressed in terms of the field $\phi^+{}_{\alpha  a}{}^i$ while equation (5.45) implies that the self-dual part of the field strength vanishes. This leads to  the  correct linearised equation for the vector field of the $D3$ brane. 
\par

We now consider the {\bf non-linear }theory using similar arguments we used for the M5 brane. The constraints of equation (5.34) can be written, using section six, in the form 
$$
\sqrt {-\gamma} \gamma ^{ \alpha \beta} \nabla _\beta x^{ \underline a}= 2
\epsilon ^{\alpha \beta \gamma_1\gamma _{2} }\nabla _\beta 
x^{\underline a \underline b_1 \underline b_2 } \nabla _{\gamma_1} x_{\underline b_{1}} \nabla _{\gamma_{2}} x_{\underline b_{2}}
\eqno(5.46)$$
\par
We also adopt the  constraint of equation (5.36) however, we must modify that of equation (5.35) as it is not reparameterisation invariant. For the non-linear theory we must take the D3 brane to have an enlarged world volume. We take it to have the coordinates $\xi^{\underline \alpha}= \{\xi^\alpha , \xi _\alpha^j \}$, in other words,  the brane moves in some of the directions with  higher level coordinates. We define 
$$
E_{\underline \alpha}{}^ A= \left( \matrix{ \nabla_\alpha x^a & \nabla_\alpha x_b^j \cr  
 \nabla^a_i x^a & \nabla^a_i x_b^j \cr }    \right) 
\eqno(5.47)$$
We note that 
$$
\delta (\nabla x^a)= -4 \Lambda _{- }{}^{ab}{}_i \nabla x_b{}^i = -4\Lambda _{- }{}^{ab}{}_i E_b^i , \quad
\delta (\nabla x_a{}^i)=0
\eqno(5.48)$$
under the transformations with parameters $\Lambda _{- }{}^{ab}{}_i$ and $\Lambda_{ab^\prime}$ provided we use the constraints of equations (5.34) and (5.36). Instead of  equation (5.35) we adopt the condition 
$$
{\cal E} _{a_1a_2} {}^i \equiv \nabla_{[a_1} x_{a_2]}{}^i 
-\nabla _b x_{[ a_1 |} {}^j\nabla _j^b x_{|a_2 ]} {}^i=0
\eqno(5.49)$$
where $\nabla_a=(s^{-1})_{a}{}^\alpha \nabla_\alpha $, $s_\alpha{}^a= \nabla_\alpha x^a$ and 
$\nabla_{j}^{b} = (E^{-1})_{j}^{b}{}^{\underline \alpha} \nabla_{\underline \alpha} $. Using  the  variations of equation (5.48) and  one finds that equation (5.49) is invariant under the local transformations provided one uses the on-shell condition (5.49) and neglects terms which contain higher level derivatives. As we mentioned for the M5 brane we are finding equations of motion to lowest order in the usual derivatives, that is, they just contain derivatives with respect to the usual world volume coordinates $\xi^\alpha$. 
In doing this we must add terms in the equation being varied which contain derivatives with respect to the coordinates $\xi_\alpha^j$. In view of the limited nature of the calculations for the non-linear theory the proposal for the non-linear theory given here must be regarded as suggestion rather than a proven result. 
\par
Neglecting all the terms with higher level derivatives the equations for the non-linear theory are those of equation (5.46) and the equation of motion for the gauge field is given by  
$$
\nabla_{[ a_1 } x_{a_2]}{}^i  = {1\over 2}\epsilon_{a_1a_2 }{}^{b_1b_2} \epsilon^{ij}  \nabla_{b_1 }  x_{b_2}{}_j
\eqno(5.50)$$

\medskip
{\bf 6 Branes in seven dimensions }
\medskip
The seven dimensional theory emerges when we decompose $E_{11}$ into 
$GL(7)\otimes Sl(5)$ which is the algebra that emerges when we 
delete node seven in the $E_{11}$ Dynkin diagram.  The generators in this decomposition at low level are easily computed using the Nutma programme SimpLie [30] and are given by 
$$ K^{\underline a}{}_{\underline b}, \ \ R^M{}_N; \ \ R^{\underline aMN} ; \ \ R^{\underline a_1\underline a_2}{}_M; \ \ R^{\underline a_1\underline a_2\underline a_3M}; \ \ R^{\underline a_1\ldots \underline a_4}{}_{MN} ; \ \ R^{\underline a_1\ldots \underline a_5M}{}_N, 
$$
$$
  R^{\underline a_1\ldots \underline a_4,\underline b}, \ \ R^{\underline a_1\ldots \underline a_6}{}_{MN,P},
; \ \ R^{\underline a_1\ldots \underline a_6}{}^{(MN)},  \ \ R^{\underline a_1\ldots \underline a_5,\underline b}{}^{ MN} , \ldots \eqno(6.1.1)$$
where $a,b,\ldots = 0,1,\ldots ,6$ and $M,N,\ldots = 1,2,3,4 ,5$. 
The generators in  $I_c(E_{11})$ algebra are  
$$ 
J^{\underline a}{}_{\underline b }= K^{\underline a}{}_{\underline b} - K^{\underline b}{}_{\underline a,} \ \ S^M{}_N = R^M{}_N - R^N{}_M,
\ \ 
S^{\underline aMN} = R^{\underline aMN} - R_{\underline aMN},  
$$
$$
S^{\underline a_1\underline a_2}{}_M = R^{\underline a_1\underline a_2}{}_M + R_{\underline a_1\underline a_2}{}^M,\ \  S^{\underline a_1\underline a_2\underline a_3M} = R^{\underline a_1\underline a_2\underline a_3M} - R_{\underline a_1\underline a_2\underline a_3M}, \ldots 
\eqno(6.1.2)$$
While the elements of the vector representation are 
$$ P_{\underline a}; \ \ Z^{MN} ; \ \ Z^{\underline a}{}_M; \ \ Z^{\underline a_1\underline a_2M}; \ \ Z^{\underline a_1\underline a_2\underline a_3}{}_{MN} ; \ \ Z^{\underline a_1\underline a_2\underline a_3,\underline b}, \ \ Z^{\underline a_1\ldots \underline a_4} , \ \ Z^{\underline a_1\ldots \underline a_4 M}{}_{N} , 
$$
$$
 \ \ Z^{\underline a_1\ldots \underline a_5 MN}, \ \ Z^{\underline a_1\ldots \underline a_5 (MN)} , \ \ Z^{\underline a_1\ldots \underline a_5}{}_{MN,P}, \ \ Z^{\underline a_1\ldots \underline a_4,\underline b MN}, \ldots   
\eqno(6.1.3)$$
The $E_{11}\otimes_s l_1 $ algebra in this decomposition has been worked out at low levels by Michaella Pettit and the author and will be given elsewhere [33]. 
\par
The non-linear realisation $I_c(E_{11})\otimes_s l_1$ is constructed from $g= g_L g_h$ where 
$$
g_l=exp (x^{\underline a} P_{\underline a}+ x_{MN} Z^{MN} +
x_{\underline a}{}^M Z^{\underline a}{}_M +x_{\underline a_1\underline a_2M}  Z^{\underline a_1\underline a_2M} + x_{\underline a_1\underline a_2\underline a_3}{}^{MN}  Z^{\underline a_1\underline a_2\underline a_3}{}_{MN} 
$$
$$+
x_{\underline a_1\underline a_2\underline a_3,\underline b} Z^{\underline a_1\underline a_2\underline a_3,\underline b} + x_{\underline a_1\ldots \underline a_4}  Z^{\underline a_1\ldots \underline a_4} +
x_{\underline a_1\ldots \underline a_4 M}{}^{N} Z^{\underline a_1\ldots \underline a_4 M}{}_{N}+ \ldots )
\eqno(6.1.4)$$
We will initially consider the most general group element $g_h \in I_c(E_{11})$, that is, before we have used the local subalgebra ${\cal H}$ to set some  of the fields in $g_h$   to zero. Thus we take 
$$
g_h= exp(\phi_{\underline a}{}^{\underline b} J^{\underline a}{}_{\underline b}+ \phi_M{}^N S^M{}_N +
\phi_{\underline aMN} S^{\underline aMN} + \phi_{\underline a_1\underline a_2}{}^M S^{\underline a_1\underline a_2}{}_M +\phi_{\underline a_1\underline a_2\underline a_3M}S^{\underline a_1\underline a_2\underline a_3M}+\ldots )
\eqno(6.1.5)$$
\par
The Cartan forms which belong to the vector representation  of the $E_{11}$ algebra  can be written in  the form 
$$
{\cal V}_l= \nabla x^{ \underline a} P_{\underline a} + \nabla x_{PQ} Z^{PQ}+
 \nabla x_{\underline a}{}^M Z^{\underline a}{}_M +  \nabla x_{\underline a_1\underline a_2M} Z^{\underline a_1\underline a_2M} + \nabla x_{\underline a_1\underline a_2\underline a_3}{}^{MN} Z^{\underline a_1\underline a_2\underline a_3}{}_{MN} 
$$
$$
+ \nabla x_{\underline a_1\underline a_2\underline a_3,b} Z^{\underline a_1\underline a_2\underline a_3,\underline b} +  \nabla x_{\underline a_1\ldots \underline a_4}  Z^{\underline a_1\ldots \underline a_4} + \nabla x_{\underline a_1\ldots \underline a_4 M}{}^{N}  Z^{\underline a_1\ldots \underline a_4 M}{}_{N} +\ldots 
\eqno(6.1.6)$$

Using equation (2.9),  the group element of equation (6.1.5) and the results of reference [33],  we find that the Cartan forms are given at low levels by 
$$
\nabla x^{\underline a} = \partial x^{\underline a}+ 2\partial x^{\underline b}\phi_{\underline b}{}^{\underline a}+ 2 \partial x _{MN} \phi^{\underline a MN}-2 \partial x^{\underline bS}\phi_{\underline b\underline aS}-12 \phi_{\underline a \underline b_1\underline b_2 M} \partial x ^{\underline b_1\underline b_2M}+\ldots ,  
$$
$$
\nabla x_{MN}= \partial  x_{MN}-2 \partial x_{S[M} \phi ^S{}_{N]}
-\partial x ^{\underline b} \phi_{\underline bMN} 
-{1\over 2} \partial x^{\underline a P}{}^{}\phi _{\underline a}^{RS}\epsilon _{RSPMN}
+2 \phi_{\underline a_1\underline a_2 [ M} \partial x^{\underline a_1\underline a_2 }{}_{N]}+\ldots , 
$$
$$
\nabla x_{\underline a}^P= \partial x_{\underline a}^P+ 2 \partial x_{\underline b}^P\phi^{\underline b}{}_{\underline a}+\partial x_a^Q \phi_Q{}^P 
+\phi_{\underline a}{}^{MN}\partial x^{RQ} \epsilon _{MNRQP} +4\phi _{\underline bMP} \partial x_{\underline a}{}^{\underline bM}+2 \phi_{\underline a\underline bP}\partial x ^{\underline b} +\ldots 
$$
$$
\nabla x_{\underline a_1\underline a_2 P}=\partial x_{\underline a_1\underline a_2 P}+ 4 \partial x ^{\underline b}{}_{ [\underline a_2 |P}\phi_{\underline b | \underline a_1 ]} +\partial x_{\underline a_1\underline a_2 Q} \phi^{Q}{}_{P}
-2 \phi _{[ \underline a_1 | QP}\partial x_{| a_2 ]}{}^{Q} -2 \phi_{a_1a_2 R} \partial x^{R}{}_{P} 
$$
$$
+6 \phi_{\underline b\underline a_1\underline a_2 P} \partial x^{\underline b }+{3\over 2} \phi ^{\underline bMN} \partial x_{\underline b\underline a_1\underline a_2 }^{RS}\epsilon_{MNRSP}+\ldots 
$$
$$
\nabla x_{\underline a_1\underline a_2\underline a_3}{}^{PQ} =
\partial x_{\underline a_1\underline a_2\underline a_3}{}^{PQ}+  6 \partial x_{\underline b [\underline a_1\underline a_2 |}{}^{ PQ} \phi ^{\underline b}{}_{| \underline a_3]} +2 \partial x_{a_1a_2a_3 S}{}^{[ Q|}\phi ^{S| P ]} -\phi _{[ a_1 | MN} \partial x_{|a_2a_3 ]T} \epsilon^{MNTPQ} 
$$
$$
-2 \phi _{[ \underline a_1\underline a_2}{}^{[ P} \partial x_{ \underline a_1] }^{Q ]} +2 \phi _{\underline a_1\underline a_2\underline a_3 M}\partial x _{RS} \epsilon^{MRSPQ}-8 \phi_{\underline b \underline a_1\underline a_2\underline a_3}{}^{PQ} \partial x^{\underline b}+\ldots 
\eqno(6.1.7)$$
where we are using form notation for the derivatives. These expressions   applies to all branes in seven dimensions once we restrict the $\phi$ fields using the local subalgebra ${\cal H}$ corresponding to the brane being considered.   
\par
Under a local transformation $I_c(E_{11})$ the Cartan forms transform as follows 
$$
\delta (\nabla x^{\underline a} )= 2\nabla x^{\underline b}\Lambda_{\underline b}{}^{\underline a}+ 2 \nabla x _{MN} \Lambda^{\underline a MN}-2 \nabla x^{\underline bS}\Lambda_{\underline b\underline aS}-12 \Lambda_{\underline a \underline b_1\underline b_2 M} \nabla x ^{\underline b_1\underline b_2M},  
$$
$$
\delta (\nabla x_{MN}) = -2 \nabla x_{S[M} \Lambda ^S{}_{N]}
-\nabla x ^{\underline b} \Lambda_{\underline bMN} 
-{1\over 2} \nabla x^{\underline a P}{}^{}\Lambda _{\underline a}^{RS}\epsilon _{RSPMN}
+2 \Lambda_{\underline a_1\underline a_2 [ M} \nabla x^{\underline a_1\underline a_2 }{}_{N]}, 
$$
$$
\delta (\nabla x_{\underline a}^P)=  2 \nabla x_{\underline b}^P\Lambda^{\underline b}{}_{\underline a}+\nabla x_a^Q \Lambda_Q{}^P 
+\Lambda_{\underline a}{}^{MN}\nabla x^{RQ} \epsilon _{MNRQP} +4\Lambda _{\underline bMP} \nabla x_{\underline a}{}^{\underline bM}+2 \Lambda_{\underline a\underline bP}\nabla x ^{\underline b} 
$$
$$
\delta (\nabla x_{\underline a_1\underline a_2 P} )= 4 \nabla x ^{\underline b}{}_{ [\underline a_2 |P}\Lambda_{\underline b |\underline a_1 ]} +\nabla x_{\underline a_1\underline a_2 Q} \Lambda^{Q}{}_{P}
-2 \Lambda _{[ \underline a_1 | QP}\nabla x_{| a_2 ]}{}^{Q} -2 \Lambda_{a_1a_2 R} \nabla x^{R}{}_{P} 
$$
$$
+6 \Lambda_{\underline b\underline a_1\underline a_2 P} \nabla x^{\underline b }+{3\over 2} \Lambda ^{\underline bMN} \nabla x_{\underline b\underline a_1\underline a_2 }^{RS}\epsilon_{MNRSP}
$$
$$
\delta (\nabla x_{\underline a_1\underline a_2\underline a_3}{}^{PQ}) = 6 \nabla x_{\underline b [\underline a_1\underline a_2 |}{}^{ PQ} \Lambda ^{\underline b}{}_{| \underline a_3]} +2 \nabla x_{a_1a_2a_3 S}{}^{[ Q|}\Lambda ^{S| P ]} -\Lambda _{[ a_1 | MN} \nabla x_{|a_2a_3 ]T} \epsilon^{MNTPQ} 
$$
$$
-2 \Lambda _{[ \underline a_1\underline a_2}{}^{[ P} \nabla x_{ \underline a_3] }^{Q ]} +2 \Lambda _{\underline a_1\underline a_2\underline a_3 M}\nabla x _{RS} \epsilon^{MRSPQ}-8 \Lambda_{\underline b \underline a_1\underline a_2\underline a_3}{}^{PQ} \nabla x^{\underline b}
\eqno(6.1.8)$$
The above transformations belong to $I_c(E_{11})$, but the actual local transformations belong to the subalgebra ${\cal H}$ and they can be obtained from the above by restricting the parameters $\Lambda$. This will provide a quick way of finding the local transformations for the different branes. The reader will notice that equations (6.1.7) and (6.1.8) are closely related as becomes  clear once one examines equations (2.8) and (2.15). 
\medskip
{\bf 6.2 The one brane}
\medskip
The one brane has a two dimensional world volume and so we take $a,b,\ldots =0,1$ and $a^\prime ,b^\prime ,\ldots =2,\ldots ,6$. The one brane charge $<Z_M^a>$ can be chosen  to be of the form  $<Z_M^a>=q_mZ^a$ where $q_Mq_M=1$. The different choices of the parameter $q_M$ allow us to turn on different charges using the same formalism. No matter what choice of $q_M$ we take it  will break the SO(5) symmetry to SO(4) which will belong to the local subalgebra ${\cal H}$. We also introduce 
an orthogonal set of parameters $q_M^i, \ i=1,2,3,4$ which obey $q_M^iq_M^j=\delta^{ij}$  and $q_Mq_M^i=0$. Using these we can write the preserved SO(4) generators as $S^i{}_j= J^M{}_N q_M^i q^N_j$ where we use summation convention on the SO(5) indices. 
\par
Examining the group element $g_l$ of equation (6.1.4) we conclude that the coordinate corresponding to the active charge is $y_a \equiv x_a^Mq_M$.  We  introduce the orthogonal  coordinates as follows  
$$
y_a \equiv x_a^M q_M ,\ \ y_a^i= x_a^M q_M^i, \ \  {\rm or \ equivalently }\ \ 
x_a^M = q^My_a + q^M_i y^i_a
\eqno(6.2.1)$$
The coordinates $x_a^M$ belong to the 5 of SO(5) and the above decomposes them into the $4+1$ of SO(4). 
The coordinates $x_{MN}$ belong to the 10 of SO(5) which decomposes into the $6+4$ of SO(4) as follows 
$$  
x^{ij}= q_M^i q_N^j x^{MN}, \ x^i= q_M^i q_N x^{MN}
\ \ ,{\rm or \ equivalently },\ \ x_{MN}=q_M^iq_N^j x_{ij} +(q_M^iq_N- q_N^iq_M) x_i
\eqno(6.2.2)$$
We adopt analogous decompositions for all the objects in the 5 and 10 representations of SO(5), for example 
$$
S^{a_1a_2} {}^{M} = q^M S^{a_1a_2} + q^M_i  S^{a_1a_2} {}^{i}
\eqno(6.2.3)$$
\par
We will choose our local subalgebra ${\cal H}$ to contain the generators 
$$
{\cal H}= \{ J_{ab},\  J_{a^\prime b^\prime}, \ S^i{}_J ,\ L_{ab^\prime} , \ S_{-aij}, \ S, S_{a^\prime ij}, \ \ldots \}
\eqno(6.2.4)$$\
where 
$$
L_{ab^\prime }= J_{ab^\prime}+ \epsilon_{a }{}^{c} S_{c b^\prime} ,\ 
S ={1\over 2} \epsilon_{a_1a_2} S^{a_1a_2} 
\eqno(6.2.5$$
$$
S_{\pm aij}= {1\over 2} (S_{aij}\pm \epsilon_{ab} \epsilon_{ijkl} S^{bkl})
\eqno(6.2.6)$$
The generators of the local subalgebra ${\cal H}$ obey the commutators 
$$
[L_{ab^\prime }, L_{cd^\prime} ] =0 ,\ [S_{- aij}, S_{-  bkl}] =0 ,\ 
[S , L_{ab^\prime} ]= -L_{ab^\prime} , \ [S , S_{- aij}]= - S_{- aij},  
$$
$$
[ L_{ab^\prime}  , S_{- bij}]=0, \ldots 
\eqno(6.2.7)$$
as well as the usual commutators with the Lorentz generators $J_{ab}$ and  $J_{a^\prime b^\prime}$. One finds that the commutator of $S_{- aij}$ and $S^{bk}$ generates $S^l$ which is not in ${\cal H}$ and so we may conclude that $S^{bk}$ is also not in ${\cal H}$. A similar argument implies that $S_{a_1a_2 k} $ is also not in ${\cal H}$. 
\par
Under the transformations of the local subalgebra generated by $L_{ab^\prime}$ and $S_{-aij}$ we find that 
$$
\delta (\nabla x^a )= -2\nabla y^{b^\prime}\tilde \Lambda _{b^\prime}{}^{ a} +2\nabla x_{ij} \Lambda_{+aij} +2 \nabla x^{b^\prime} \Lambda_{b^\prime}{}^{a}, \ 
$$
$$
\delta (\nabla y_a )= -2\nabla y^{b^\prime} \Lambda _{b^\prime a}+\epsilon _{ijkl} \Lambda ^{+aij} \nabla_{kl} +2 \nabla x^{b^\prime} \tilde \Lambda _{b^\prime a}
$$
$$
\delta (\nabla x^{a^\prime} )= 2(\nabla x_b -\epsilon_{bc} \nabla y^c)\Lambda^{b}{}^{a^\prime} ,\ 
\delta (\nabla y^{a^\prime} )= -2\epsilon_{b}{}^{d}(\nabla x_d -\epsilon_{dc} \nabla y^c)\Lambda^{b}{}^{a^\prime} ,\ 
$$
$$
\delta(\nabla x_{ij})= -(\nabla x^b -\epsilon^{bc} \nabla y_c)\Lambda_{+b}{}_{ij} 
\eqno(6.2.8)$$
We can derive these transformations from  equations (6.1.8) provided we restrict the $\Lambda$ parameters so that they belong to the subalgebra, that is, adopt the constraint $\Lambda _{a}{}^{b^\prime} = \epsilon _{ac }\tilde \Lambda ^{cb^\prime} $ where $\tilde \Lambda ^{cb^\prime} = \Lambda ^{cb^\prime} {}_{M} q^M$ and the  corresponding self-duality condition for  $\Lambda_{+a ij}$. 
\par
One can verify that the equations
$$
\nabla_\alpha x^a -\epsilon^{ab} \nabla _\alpha y_b =0
= \nabla_\alpha  x^{a^\prime}=\nabla_\alpha  y^{a^\prime}=\nabla _\alpha x_{ij}= 0  
\eqno(6.2.9)$$
are left invariant under the transformations of equation (6.2.8). 
\par
It is very instructive to analyse the equations (6.2.9) at the linearised level for which we  assume  static gauge $\partial_\alpha x^a= \delta _\alpha^a$ and so $\partial_\alpha y_a= -\epsilon_{\alpha a}$. 
From the expressions for the Cartan forms of equation (6.1.7) we find, without making any restrictions on the fields $\phi$,  that 
$$
\nabla_\alpha x^{a^\prime} - \epsilon_{\alpha }{}^{\beta} \nabla_\beta y^{a^\prime}= \partial_\alpha x^{a^\prime} - \epsilon_{\alpha }{}^{\beta} \partial_\beta y^{a^\prime} +\ldots ,\  
\nabla_\alpha x^{a^\prime} + \epsilon_{\alpha }{}^{\beta} \nabla_\beta y^{a^\prime}= \partial_\alpha x^{a^\prime} + \epsilon_{\alpha }{}^{\beta} \partial_\beta y^{a^\prime} +4\hat \phi_\alpha{}^{a^\prime}+ \ldots ,\  
$$
$$
\nabla _\alpha  x_{ij}= \partial_\alpha x_{ij} -4 \phi_{-\alpha ij}+\ldots 
\eqno(6.2.10)$$
The choice of local subalgebra ${\cal H}$ can be deduced from the requirement that the above expressions for $\phi$'s of equation (6.2.10) do occur in the group element $g_h$ and that their orthogonal combinations do not. Put another way it means that the orthogonal combinations occur in a  local subalgebra transformation $h$ and so can be removed from $g_h$. 
Since all the objects that occur in equation (6.2.10) vanish we find that half of the equations  are inverse Higgs conditions and half  are equations of motion. 
\par
The equations of motion can, following the arguments in section six,  be written as 
$$
\sqrt {-\gamma} \gamma ^{\alpha \beta} \nabla _\beta x^{\underline a}= 
- \epsilon ^{\alpha \beta} \nabla_\beta  y^{\underline a} , \ 
\sqrt {-\gamma} \gamma ^{\alpha \beta}\nabla_\beta  x_{ij}= -
{1\over 2}\epsilon^{\alpha \beta}  \epsilon_{ijkl}  \nabla _\beta x^{kl} 
\eqno(6.2.11)$$
It would be interesting to analyse the content of these equations in detail. It would also be interesting to study the effect of the higher level local symmetries and  the higher level conditions that they will impose. 
\par
We can count the number of bosonic degrees of freedom in this seven dimensional theory. We have $7-2=5$ degrees of freedom in $x^{a^\prime} $ and 
${4.3\over 2.2}= 3$ from $x_{ij}$ which gives us $8$ bosonic degrees of freedom. This is the number required for a half BPS brane that is maximally supersymmetric.  In making this count we have assumed that the coordinates $x^i, x_a{}^i,\dots $ do not contribute to the brane degrees of freedom.

\medskip
{\bf 6.3 The two brane}
\medskip
The two brane has a three dimensional world volume and so we take $a,b,\ldots =0,1,2$ and $a^\prime ,b^\prime ,\ldots =3,\ldots ,6$.  Following the same decompositions as in the previous sections, the rank two charge can be written as $Z^{a_1a_2 M}= q^M Z^{a_1a_2} + q^M_i Z^{a_1a_2 i}$. We can choose the  $Z^{a_1a_2}$ charge to be the one that is active so breaking SO(5) to SO(4).  We adopt similar decompositions of the generators and other objects as in the last section. 
\par
We  choose our local subalgebra ${\cal H}$ to contain the generators 
$$
{\cal H}= \{ J_{ab},\  J_{a^\prime b^\prime}, \ S^i{}_j ,\ L_{ab^\prime} , \ L_{ai}, \ S,  \ \ldots \}
\eqno(6.3.1)$$\
where 
$$
L_{ab^\prime }= J_{ab^\prime}-{1\over 4}  \epsilon_{a }{}^{e_1e_2} S_{e_1e_2 b^\prime} ,\  
L_{ai}= S_{ai}+{1\over 2}  \epsilon_{a}{}^{ e_1e_2} S_{ e_1e_2 i} ,\ \ 
S ={1\over 3!} \epsilon_{a_1a_2a_3} S^{a_1a_2a_3} 
\eqno(6.3.2)$$
These generators obey the algebra 
$$
[L_{ab^\prime }, L_{cd^\prime} ] =0 ,\ [S_{ai}, S_{ b j}] =0 ,\ 
[S , L_{ab^\prime} ]= 2 L_{ab^\prime} , \ [S , L_{a i}]= 2 L_{ai }, \ldots 
\eqno(6.2.3)$$
as well as the usual commutators with the Lorentz generators $J_{ab}$ and  $J_{a^\prime b^\prime}$.
\par
Using the transformations of equation (6.1.8) but with the restrictions 
$$
\Lambda_{ab^\prime}= 3\epsilon_{a}{}^{e_1e_2} \Lambda _{e_1e_2 b^\prime} ,\ 
\Lambda_{ai}= {1\over 2} \epsilon_{a}{}^{e_1e_2} \Lambda_{e_1e_2 i} ,\ 
\Lambda_{a_1a_2a_3}={1\over 3!} \epsilon_{a_1a_2a_3}\Lambda
\eqno(6.3.4)$$
so as to ensure that the transformation do belong to the local subalgebra,  ${\cal H}$ we find that 
$$
\delta (\nabla x^{a} )= 2\nabla x^{b^\prime}\Lambda _{b^\prime} {}^{a}-\epsilon_{ab_1b_2}\Lambda \nabla x^{b_1b_2} +4 \nabla x_i \Lambda^{ai}
-2\epsilon_{abc }\Lambda_{ci} \nabla x^{bi}+4 \epsilon_{abe} \Lambda _{ec^\prime} \nabla x^{bc^\prime}
$$
$$
\delta (\nabla x^{a^\prime} )= 2 {\cal E} ^b \Lambda _{b}{}^{a^\prime}  , \ 
\delta (\nabla x_i)= -\Lambda _{bi} {\cal E} ^b ,\ 
\delta (\nabla x_{ai}) = -2 \epsilon_{abc} {\cal E} ^b \Lambda^{c}{}_{i} +\ldots 
$$
$$
\delta (\nabla x_{a_1a_2})= 4\nabla x^{b^\prime} {}_{[ a_1} \Lambda _{a_2 ]b^\prime}- 2\Lambda_{[ a_1 | i} \nabla x_{ | a_2]}{}^{i}-2\Lambda _{a_1a_2 i} \nabla x^i +\epsilon_{a_1a_2 b} \nabla x^b
$$
$$
\delta (\nabla x_{ab^\prime})= -2\nabla x_{b^\prime}{}^{ c^\prime}\Lambda _{c^\prime a} -\Lambda_{ai} \nabla x_{b^\prime }{}^{i} +\epsilon_{acd} \Lambda^d{}_{b^\prime}{\cal E}^c
\eqno(6.3.5)$$
where ${\cal E} ^a= \nabla x^a + \epsilon^a{}_{e_1e_2}\nabla x^{e_1e_2}$. Using these variations we find that 
$$
\delta ({\cal E} ^a) = -2\Lambda {\cal E} ^a -2 \Lambda _{ac^\prime}\nabla x^{c^\prime} 
\eqno(6.3.6)$$
\par
The local symmetries are preserved if we adopt the conditions 
$$
{\cal E} ^a_\alpha= \nabla_\alpha x^{a^\prime}= 0= \nabla_\alpha x_i  
\eqno(6.3.7)$$
We now focus on  the {\bf linearised}  theory and adopt the additional conditions   
$$ 
\nabla _{[ \alpha} x_{a ]c^\prime}=0 ,\ \nabla_{[\alpha} x_{a ]i}=0
\eqno(6.3.8)$$
provided we also demand certain higher level constraints. We note that we could have taken no antisymmetry on the above conditions and this would still preserve the symmetries. We have also taken only the lowest level brane world volume coordinate to be active. 
\par
To examine the meaning of these constraints we evaluate the Cartan forms at the linearised level.  Using the lowest order conditions $\partial_\alpha x^a=\delta_\alpha^a$ and $\partial_\alpha x_{b_1b_2}= {1\over 2} \epsilon _{\alpha b_1b_2} $ and equation (7.1.7) we find taking the most general $\phi$'s that 
$$
\nabla_\alpha  x^{a^\prime} + \epsilon _{\alpha }{}^{c_1c_2} \nabla_{c_1} x_{c_2} {}^{a^\prime}=0, \ 
\nabla_\alpha  x^{a^\prime} - \epsilon _{\alpha }{}^{c_1c_2} \nabla_{c_1} x_{c_2} {}^{a^\prime}=8\Phi_{\alpha a^\prime} 
\eqno(6.3.9)$$
$$
\nabla_\alpha  x_i +{1\over 4} \epsilon _{\alpha }{}^{c_1c_2} \nabla_{c_1} x_{c_2 i} =0 ,\ \nabla_\alpha  x_i -{1\over 4} \epsilon _{\alpha }{}^{c_1c_2} \nabla_{c_1} x_{c_2 i} = -4\Phi _{\alpha i} 
\eqno(6.3.10)$$
where 
$$
\Phi_{\alpha a^\prime}= {1\over 2} (\phi_{\alpha a^\prime} -3 \epsilon_{\alpha}{}^{c_1c_2} \phi_{c_1c_2a^\prime} ) ,\ 
\Phi _{\alpha i}= {1\over 2} (\phi_{\alpha i}-{1\over 2} \epsilon_{\alpha}
{}^{c_1c_2} \phi_{c_1c_2 i}) 
\eqno(6.3.11)$$
We see that the conditions 
$\nabla_\alpha x^{a^\prime}= 0$ and  $\nabla _{[ \alpha} x_{a ]c^\prime}=0$ 
are half equations of motion and half inverse Higgs conditions. The same holds for the conditions $\nabla_{[\alpha} x_{a ]i}=0$ and  
$\nabla_\alpha x_i =0$. The number of bosonic degrees of freedom are 4 for $x^{a^\prime}$ and 4 for $x_i$ giving a total of 8 bosonic degrees of freedom. This is the correct number for a maximally supersymmetric brane in a type II theory. 
\par
We now consider the {\bf non-linear theory} and make a proposal along similar lines to that we have given for the M5 and D3 branes. We extend the world volume of the brane so that it has the coordinates $\xi^{\underline \alpha} 
=\{\xi^\alpha , \xi_\alpha ^i , \xi_{\alpha \beta^\prime}\}$. and introduce the object $E_{\underline \alpha}{}^{A} = \nabla_{\underline \alpha} x^A$. 
where $x^A = \{ x^a x_a^j, x_{ac^\prime} \}$. We adopt the constraints of equation (6.3.7) but we replace the constraints of equation (6.3.8) by 
$$
\hat \nabla _{[ a_1} x_{a_2  ]c^\prime}=0 ,\ \hat \nabla_{[a_1} x_{a_2 ]i}=0
\eqno(6.3.12)$$ 
where 
$$
\hat \nabla _a = ( (s^{-1}) _a{}^\alpha +(\nabla _j^b x_a^j - \nabla^{bd^\prime} x_a{}_{d^\prime})(s^{-1}) _b{}^\alpha )\nabla_\alpha
\eqno(6.3.13)$$
where $\nabla _A= (E^{-1})_A{}^{\underline \alpha} \nabla_{\underline \alpha}$. As before we vary so as to keep only terms with derivatives with respect to $\xi^\alpha$ and we used equation (6.35) in the form 
$$
\delta (\nabla x^{a} )= \Lambda \nabla x^{a} 
-2\epsilon_{abc }\Lambda_{ci} E^{bi}+4 \epsilon_{abe} \Lambda _{ec^\prime} E^{bc^\prime}
\eqno(6.3.14)$$
and we have used the conditions of equation (6.3.7). 
\medskip
{\bf 7. Some generic features of the brane dynamics }
\medskip
In this section we will discuss some of the generic features that emerge from formulating brane dynamics as a non-linear realisation of $E_{11}\otimes_s l_1$, as explained in section two. The dynamics of  any  brane is given in terms of the coordinates $x^A$, which are in one to one correspondence with the elements of the vector representation,  and the fields $\varphi$ arising from  $I_c(E_{11})$ as well as   background fields which arise from  the Borel subalgebra of $E_{11}$. The charge of the brane being considered is one of the charges in the vector representation and the dynamics of the brane will involve the coordinate, corresponding to this charge,   together with the coordinate $x^{\underline a}$ associated with usual spacetime translations $P_{\underline a}$,  as well as generically  other coordinates. 
\par
It  turns out that  for every element in the Borel subalgebra of $E_{11}$ there is an element in the $l_1$ representation  [31]. At low levels this relation is one to one but at higher levels there is more than one element in the $l_1$ representation for each element in the Borel subalgebra of $E_{11}$. In the non-linear realisation  every element in the Borel subalgebra of $E_{11}$ leads to a background field and the brane couples to the field associated to the charge in the vector  representation that it carries. To give a simple example, if we are considering the M2 brane its charge is $Z^{\underline a_1\underline  a_2}$, which is the second element in the vector representation in eleven dimensions,  and  the coordinates $x_{\underline a_1\underline a_2}$  and $x^{\underline a}$ will play an important role in the  dynamics of the M2 brane.  The charge $Z^{\underline a_1\underline a_2}$ corresponds in the Borel subalgebra of $E_{11}$ to the generators $R^{\underline a_1\underline a_2\underline a_3}$ and so the background field 
$A_{\underline a_1\underline a_2\underline a_3}$ to which the M2 brane couples. These were indeed the features we found when we considered the dynamics of the M2 brane in section 4.1. 
\par
We will now illustrate, in outline only,  some of the features of the  dynamics of  a  p brane whose charge has one block of totally antisymmetrised spacetime indices.   In this case the corresponding charge is of the form $Z^{\underline a_1\ldots \underline a_p \bullet }$ in the $l_1$ representation which leads to the corresponding coordinate 
$x_{\underline a_1\ldots \underline a_p \bullet }$ in the non-linear realisation where $\bullet$ represents the  internal group indices which we will suppress in what follows. We also have the  coordinate $x^{\underline a}$ associated with the spacetime translations $P_{\underline a} $. These occur in the Cartan form as follows; 
$$
{\cal V}_l= \nabla_\alpha x^{\underline a} P_{\underline a}+\ldots + \nabla_\alpha  x_{\underline a_1\ldots \underline a_p }Z^{\underline a_1\ldots \underline a_p }+\ldots 
\eqno(7.1)$$
\par
Associated with the p-form charge is an element of the Borel subalgebra of $E_{11}$ of the form $R^{\underline a_1\ldots \underline a_{p+1}}$ which obeys a commutator of the form 
$$
[R^{\underline a_1\ldots \underline a_{p+1}} , P_{\underline b} ]= 
 f_1 \delta ^{[\underline  a_1} _{b}Z^{\underline a_2\ldots \underline a_{p+1}]}
\eqno(7.2)$$
where $f_1$ is a constant whose value is known from  the $E_{11}$ algebra. Associated with the generator $R^{\underline a_1\ldots \underline a_{p+1} }$  is a corresponding element of 
$I_c(E_{11})$ of the form $S^{\underline a_1\ldots \underline a_{p+1} }$ which has the following generic commutators 
$$
[S^{\underline a_1\ldots \underline a_{p+1}} , P_{\underline b} ]= 
f_1 \delta ^{[\underline  a_1} _{b}Z^{\underline a_2\ldots \underline a_{p+1}]}, \quad
[S^{\underline a_1\ldots \underline a_{p+1}} , Z^{\underline b_1\ldots \underline b_p} ]= 
f_2 \delta ^{[\underline a_1 \ldots \underline a_p}_{\underline b_1\ldots \underline b_p }P^{ \underline a_{p+1}]}
\eqno(7.3)$$
where   $f_2$ is a  constant that is also  specified by the $E_{11}$ algebra. 
If we consider the indices on the generator $S^{\underline a_1\ldots \underline a_{p+1}}$ to take the values $a,b,\ldots = 0,1,\ldots , p$, which are those in the brane direction,  and define  $ S^{ a_1\ldots  a_{p+1}}\equiv -(p+1)! \epsilon^{ a_1\ldots  a_{p+1}} S $ then the above commutators become 
$$
[S , P_{ b} ]= 
f_1 \epsilon_{b a_1\ldots  a_{p}} Z^{ a_1\ldots  a_{p}}, \quad
[S, Z^{ b_1\ldots  b_p} ]= 
f_2 \epsilon^{ b_1 \ldots  b_p c}P_{ c}
\eqno(7.4)$$
\par
It will turn out that the  generator $S$ belongs to the local subalgebra ${\cal H}$ and so it leads to a local symmetry of the non-linear realisation whose action is  given in equation (2.15). As a result the Cartan forms of equation (6.1) transform as 
$$
\delta {\cal V}_l = -[\Lambda S , {\cal V}_l]
\eqno(7.5)$$
and using the commutators of equation (7.3) we find that 
$$
\delta (\nabla _\alpha x^{ a} )= f_2 \Lambda \epsilon_{ b_1\ldots  b_{p} a} \nabla _\alpha x^{ b_1\ldots  b_{p}} , \quad 
\delta ( \nabla _\alpha x_{ a_1\ldots  a_p} )= \Lambda f_1 \nabla _\alpha x^{  b}
\epsilon_{ b  a_1\ldots  a_p}
\eqno(7.6)$$
\par
An equation that is invariant under $SO(p+1)\otimes SO(11-p-1)$ Lorentz transformations, rigid $E_{11}$ transformations, world volume diffeomorphism and the local transformations of equation (7.6) is given by  
is 
$$
\nabla _\alpha x^a= -e_1 \epsilon^{a b_1 \ldots b_p} \nabla _\alpha x_{b_1\ldots b_p}\ \ {\rm equivalent\  to }\ \ \nabla_\alpha x^{a_1\ldots a_p}= e_2\epsilon^{b a_1\ldots a_p} \nabla _\alpha x_b 
\eqno(7.7)$$
provided  the constant $e_2={1\over e_1 p!}$ and $e_1$  obeys the condition $e^2_1= -(-1)^p f_2  f_1^{-1}( p!)^{-1}$. 
By varying this equation under the other transformations of the local subalgebra one can find at least some of the other equations of motion of the brane dynamics. Clearly one  has to verify that the full set of equations of is invariant under all the symmetries of the non-linear realisation. 
\par
We can rewrite the second equation in (7.7) in the form 
$$
(s^{-1} )_b{}^\alpha\nabla_\alpha x^{a_1\ldots a_p}= e_2\epsilon^{b a_1\ldots a_p} 
\eqno(7.8)$$
where $s_\alpha ^b\equiv \nabla _\alpha x^b $. Using the identity 
$$
\epsilon^{\alpha_1 \beta_1 \ldots \beta _p} s_{\alpha_1}^b s_{\beta_1}^{a_1}\ldots  s_{\beta_p}^{a_p}= (\det s)\epsilon ^{ba_1\ldots a_p}
\eqno(7.9)$$
in equation (7.8) and multiplying by a further factor of $s^{-1}$ we find the  equation 
$$
\sqrt {-\gamma} \gamma ^{\alpha \beta} \nabla _\beta x^{a_1\ldots a_p} =e_2
\epsilon ^{ \alpha\gamma_1 \ldots \gamma _{p-1} }
\nabla _{\gamma_1} x_{b_{1}}\ldots \nabla _{\gamma_p} x_{b_{p}}
\eqno(7.10)$$
where 
$$
\gamma _{\alpha \beta} \equiv  (s\eta s^T)= \nabla_\alpha x^a \eta_{ab} \nabla _\beta x^b \quad {\rm and }\quad \gamma= det \gamma _{\alpha \beta} =-(\det s)^2
\eqno(7.11)$$ 
\par

Using an identity similar to that of equation (7.9) we find that 
$$
\sqrt {-\gamma} \gamma ^{\alpha \beta} \nabla _\beta x^a= e_1 
\epsilon ^{\alpha \beta \gamma_1\ldots \gamma _{p-1} }\nabla _\beta 
x^{a b_1\ldots b_{p-1} } \nabla _{\gamma_1} X_{b_{1}}\ldots \nabla _{\gamma_{p-1}} X_{b_{p-1}}
\eqno(7.12)$$
\par
The steps leading to equations (7.10) and (7.12) from equation (7.7) can be reversed and so both of these latter equations are equivalent to equation (7.7) and so also to each other. These equations are  not the usual brane equations for a p-brane,  but as we have explained for the branes we have studied in this paper  we found that $\nabla_\alpha x^{a^\prime}=0$ and this allows us to extend the range on the indices in equation (7.12) from $a, \ldots $ to $\underline a , \dots $. One can then show  that  the $\varphi$ fields then disappear from the equations and one finds the familiar brane dynamics at least for the coordinates $x^{\underline a}$. 
\par
In addition to the first order duality equations for the coordinates 
$x^{\underline a}$ and $ x_{\underline a_1\ldots \underline a_p}$ there will be other dynamical equations involving  the other coordinates of the non-linear realisation, that is, the vector representation. Since the full set of equations must be invariant under the symmetries of the non-linear realisation, which preserve the number of derivatives,  the other equations   must also be first order in derivatives and will also be duality equations that describe the world volume fields living on the world volume of the brane. This pattern is borne out for the branes  we have studied in this paper. 
\par
We will now discuss how the above p-brane couples to the background fields. As we noted above the brane has a charge $Z^{\underline a_1\ldots \underline a_p}$ in the vector representation,  associated with  the coordinates $x_{\underline a_1\ldots \underline a_p}$, and related to the generators $R^{\underline a_1\ldots \underline a_{p+1}}$ in the Borel subalgebra of $E_{11}$ which are  in turn associated with the $E_{11}$ background fields $A_{\underline a_1\ldots \underline a_{p+1}}$. 
Looking at our definition of the vielbein given below equation (2.9)  and 
that the group element $g_E$ contains the factor $ e^{A_{\underline a_1\ldots \underline a_{p+1}}R^{\underline a_1\ldots \underline a_{p+1}}}$, we conclude that the vielbein  has the component 
$$
E_{\underline \mu}{}^{\underline a_1\ldots \underline a_{p}}= - f_1 A_{\underline \mu\underline a_1\ldots \underline a_{p}}+\ldots 
\eqno(7.13)$$ 
and as a result 
$$
\nabla^B_\alpha x_{\underline a_1\ldots \underline a_{p}}
= \partial_\alpha x_{\underline a_1\ldots \underline a_{p}} -f_1\partial_\alpha x^{\underline \mu} A_{\underline \mu\underline a_1\ldots \underline a_{p}}+\ldots 
\eqno(7.14)$$
We observe that this Cartan form, when its indices are in the brane world volume directions,   and are totally antisymmetrised,  is invariant under the gauge transformations 
$$
\delta  x_{ a_1\ldots  a_{p}} = f_1\Lambda _{ a_1 a_2\ldots  a_{p}  } , \quad 
\delta  A_{ a_1\ldots a_{p+1}}= \partial _{ [   a_1}\Lambda _{a_2\ldots  a_{p+1}]} 
\eqno(7.15)$$ 
provided  we adopt static gauge.   
\par
While the above parts of the brane dynamics and background coupling were worked out for a simple branes,  they illustrate the general pattern that applies to all types of branes since the above results  were  derived from features of the $E_{11}\otimes l_1$ non-linear realisation that apply to all branes, that is,  the relations between the brane charges in the $l_1$ representation and the generators in the Borel subalgebra which in turn  implies the connections between the coordinates and the fields. 
\par
We have seen that we have a different choice of local subalgebra ${\cal H}$ 
for each brane. This is to be expected as different branes break different parts of the $E_{11}$ symmetry, which, of course,  includes the Lorentz symmetry.  We observe that the local subalgebra ${\cal H}$  has the property that  the generators of the vector representation decomposes under the action of the local subalgebra so as to contain a subrepresentation. In the cases we have studied the first element in this subrepresentation was 
$$
P_a + e_2 \epsilon _{a}{}^{b_1\ldots b_p} Z_{b_1\ldots b_p}
\eqno(7.16)$$
We recognise this as the constant part of the Cartan form ${\cal V}_l$ in static gauge for which, using equation (7.7),  is given by 
$$
\partial _\alpha x^a=\delta_\alpha^a , \ \   \partial _\alpha x^{a_1\ldots a_p}= e_2\epsilon^{\alpha  a_1\ldots a_p} 
\eqno(7.17)$$
The result means that the Cartan form ${\cal V}_l$ has no constant term in its transformation under a local transformation. 
\par

It would be good to understand in a systematic way the correspondence between a chosen brane and its  local subalgebra.  One might imagine that that once one is given the brane charge in the vector representation one can deduce the corresponding local subalgebra by a well defined procedure.   Given the choice of local subalgebra the dynamics of the brane are largely determined and so such an understanding may be important for understanding the general properties of branes. 
\medskip
{\bf 8. Conclusion }
\medskip
In this paper we have further developed the theory of the non-linear of  $E_{11}\otimes_s l_1$ to find brane dynamics. The brane moves through the spacetime which is automatically contained in the non-linear realisation and has coordinates in the vector representation. These contain the usual embedding coordinates as well as the world volume fields. The equations which emerge from the non-linear realisation 
are first order in derivatives and can be thought of as a set of duality equations. The dynamics for each brane is invariant under the full $E_{11}$ symmetries,  however, which parts are linearly realised and which non-linearly realised varies from brane to brane. This difference is reflected in the different choice of local subalgebra. 
\par
We have used this theory  to find the dynamics of the strings in IIA and IIB theory,  the M2  and M5 branes, the D3 brane in IIB as well as that for  the  one and two branes in seven dimensions. The construction of the linearised dynamics for all these branes is straightforward as well as for the non-linear theory for branes with world volume fields that do not carry Lorentz indices. 
However, our construction  for the non-linear theory   for the branes that contain world volume fields which carry Lorentz indices should be regarded as only a proposal. While some part of the dynamics of such branes is clear a problem arises with ensuring the gauge symmetries of such world volume fields.  The proposed solution is to take 
 the brane to depend on an enlarged world volume. This is natural in the sense that the brane moves  not just in the usual Minkowski spacetime but also in the directions of the higher level coordinates that are automatically encoded in the E theory approach. These new brane coordinates ensure the gauge symmetry of these world volume fields in a way that parallels  the mechanism that is know to be present when the non-linear realisation is applied to find the field theories, that is,  extensions of the maximal supergravities. However, our calculations in the brane case are limited and it would be good to extend them further and gain further evidence for this approach. 
\par
There should be no obstacle to using the non-linear realisation to construct the  maximally supersymmetric brane dynamics at the linearised level in   all dimensions. As we have mentioned, one will finds a system of equations that are first order in derivatives acting on the coordinates of the vector representation. This implies that there is such a dual formulation  for all branes. We note that the usual formulation of brane dynamics which follows from an action involves a derivative acting on the fields. As such their equations of motion will involve a derivative acting on an expression and one can find a  formulation by integrating the  equations of motion to find equations that are first order in derivatives and can be viewed as duality relation. Indeed one could derive these equations, independent of the E theory approach, where the dynamics is known. 
\par
In the brane dynamics we have constructed we only computed the dynamics for the lowest  level coordinates and it remains to be found if there is further information in the higher level coordinates. If this were the case it could lead to features of brane dynamics not so far encountered. A similar remark applies to the use of the enlarged world volume. 
\par
In this paper we have not considered the construction of the Wess-Zumino terms  and it would be interesting to consider this term from the viewpoint of the non-linear realisation. A discussion of this term using certain  equations derived from E theory can be found in [29]. 
\par
A different approach to brane dynamics based on current algebra was given in reference [32] and it would be interesting to understand the connection to the viewpoint of this paper. 
\par
As we we have explained in the introduction,  $E_{11}$, through its  vector representation,  predicts the existence of an infinite number of new branes almost all of which are exotic branes. The non-linear realisation discussed in this paper has the potential to give the dynamics of all these new objects. 
\par
The infinite number of new branes that $E_{11}$ predicts in the vector representation could play an important role in understanding the entropy of black holes in the sense that they can be building blocks for a microscopic description of the entropy. Indeed one can wonder if the entropy is hidden in $E_{11}$ and the vector representation. 
\medskip
{\bf {Acknowledgements}}
\medskip
I wish to thank Michaella Pettit for help at an early stage of this work. The results found with Michaella Pettit  in reference [33], namely the Cartan involution invariant subalgebras in the IIB and the seven dimensional theories,  were crucial for some of the calculations in this paper. I also wish to thank Paul Cook for discussions on brane dynamics.   We wish to thank the SFTC for support from Consolidated grant number ST/J002798/1. 

\medskip
{\bf {References}}
\medskip
\item{[1]} P. West, {\it $E_{11}$ and M Theory}, Class. Quant.  
Grav.  {\bf 18}, (2001) 4443, hep-th/ 0104081. 
\item{[2]} P. West, {\it $E_{11}$, SL(32) and Central Charges},
Phys. Lett. {\bf B 575} (2003) 333-342,  hep-th/0307098. 
\item{[3]} A. Tumanov and P. West, {\it E11 must be a symmetry of strings and branes },  arXiv:1512.01644. 
\item{[4]} A. Tumanov and P. West, {\it E11 in 11D}, Phys.Lett. B758 (2016) 278, arXiv:1601.03974. 
\item{[5]} P. West, {\it Introduction to Strings and Branes}, Cambridge University Press, 2012. 

\item{[6]}  A. Kleinschmidt and P. West, {\it  Representations of G+++
and the role of space-time},  JHEP 0402 (2004) 033,  hep-th/0312247.
\item{[7]} P. Cook and P. West, {\it Charge multiplets and masses
for E(11)}, ÊJHEP {\bf 11} (2008) 091, arXiv:0805.4451.
\item{[8]} P. West,  {\it $E_{11}$ origin of Brane charges and U-duality
multiplets}, JHEP 0408 (2004) 052, hep-th/0406150. 
\item{[9]} P. West, {\it Brane dynamics, central charges and
$E_{11}$}, JHEP 0503 (2005) 077, hep-th/0412336.
\item{[10]} 
 S. Elitzur, A. Giveon, D. Kutasov and E.  Rabinovici,  {\it
Algebraic aspects of matrix theory on $T^d$ }, {hep-th/9707217}; 
  N. Obers,  B. Pioline and E.  Rabinovici, {\it M-theory and
U-duality on $T^d$ with gauge backgrounds}, { hep-th/9712084}; 
 N. Obers and B. Pioline,~ {\it U-duality and  
M-theory, an
algebraic approach}~, { hep-th/9812139}; 
 N. Obers and B. Pioline,~ {\it U-duality and  
M-theory}, {hep-th/9809039}. 
\item{[11]} P. Cook, {\it Exotic E11 branes as composite gravitational solutions}, Class.Quant.Grav.26 (2009) 235023, arXiv:0908.0485.
\item{[12]} P. Cook, {\it Bound States of String Theory and Beyond } ,  JHEP 1203 (2012) 028,   arXiv:1109.6595. 
\item{[13]} J. Hughes, J. Polchinski, Nucl. Phys. {\bf B 278} (1986)
147;  J. Hughes, J. Liu, J. Polchinski, Phys. Lett. {\bf B 180}
(1986) 370;  J. Bagger, A. Galperin, Phys. Lett. {\bf B 336}
(1994) 25;  Phys. Rev. {\bf D 55} (1997)
1091;  Phys. Lett. {\bf B 412} (1997)
296; 
F. Gonzalez-Rey, I.Y. Park, M. Ro\v{c}ek, Nucl. Phys. {\bf B 544}
(1999) 243;
 M. Ro\v{c}ek, A. Tseytlin, Phys. Rev. {\bf D 59} (1999)
106001. S. Bellucci, E. Ivanov, S. Krivonos, Phys. Lett. {\bf B
460} (1999) 348; E. Ivanov, S. Krivonos, Phys. Lett. {\bf B 453}
(1999) 237; 
S. Belucci, E. Ivanov and S. Krivonos, { Partial
breaking $N=4$ to $N=2$: hypermultiplet as a Goldstone superfield\/},
hep--th/9809190.\
S. Belucci, E. Ivanov and S. Krivonos, { Partial breaking of
$N=1$, $D=10$ supersymmetry}, hep--th/9811244;\
F. Gonzalez--Rey, I. Y. Park and M. Ro\~cek, { Nucl. Phys.} {\bf
B544}  (1999) 243;\
M. Ro\~cek and A. Tseytlin, { Phys. Rev.} {\bf D59} (1999)
106001;\  E. Ivanov and S. Krivonos,
{ Phys. Lett.\/} {\bf B453} (1999) 237;\
S. V. Ketov, { Mod. Phys. Lett.\/} {\bf A14} (1999) 501;\
{ Born-Infeld-Goldstone superfield actions for gauge-fixed D-5- and
D-3-branes in 6d\/}, hep-th/9812051; E. Ivanov, S. Krivonos, {\it N=1 D=4 supermembrane in the coset approach},  Phys.Lett. B453 (1999) 237, hep-th/9901003; Gauntlett, K. Itoh, and P. Townsend, {\it Superparticle with extrinsic curvature}, Phys.Lett. B238, (1990) 65; J. P. Gauntlett, J. Gomis and P. K. Townsend, {\it Particle Actions As Wess-Zumino
Terms For Space-Time (Super) Symmetry Groups}, Phys. Lett. B 249 (1990) 255. 
\item{[14]} P. West, {\it Automorphisms, Non-linear Realizations and Branes },  JHEP0002:024,2000,  arXiv:hep-th/0001138. 
\item{[15]} P. West, {\it Hidden Superconformal Symmetry in M Theory}, 
 JHEP 0008:007,2000,  arXiv:hep-th/0005270.  
\item{[16]}  P. West, {\it Generalised space-time and duality}, 
Phys.Lett.B693 (2010) 373, arXiv:1006.0893 
\item{[17]} M. Duff, {\it Duality Rotations In String Theory},
  Nucl.\ Phys.\  B {\bf 335} (1990) 610. 
\item{[18]} A. Tseytlin, {\it Duality Symmetric Formulation Of String World
Sheet Dynamics}, Phys.Lett. {\bf B242} (1990) 163, {\it Duality Symmetric
Closed String Theory And Interacting Chiral Scalars}, Nucl.\ Phys.\ B {\bf
350}, 395 (1991). 
\item{[19]} M. Duff and J. Lu, {\it Duality rotations in
membrane theory},  Nucl. Phys. {\bf B347} (1990) 394. 
\item{[20]} M. Duff, J. Lu, R. Percacci, C. Pope, H. Samtleben and  E. Sezgin. {\it Membrane Duality Revisited },  arXiv:1509.02915. 
\item{[21]} P. West, {\it E11, generalised space-time and IIA string
theory}, 
 Phys.Lett.B696 (2011) 403-409,   arXiv:1009.2624.
\item{[22]} W. Siegel, {\it Two vielbein formalism for string inspired axionic gravity},   Phys.Rev. D47 (1993) 5453,  hep-th/9302036; 
\item{[23]} W. Siegel,{\it Superspace duality in low-energy superstrings}, Phys.Rev. D48 (1993) 2826-2837, hep-th/9305073; 
{\it Manifest duality in low-energy superstrings},  
In *Berkeley 1993, Proceedings, Strings '93* 353,  hep-th/9308133. 
\item{[24]} A. Rocen and P. West,  {\it E11, generalised space-time and
IIA string theory;  the R-R sector}, in {\it Strings, Gauge fields and the Geometry 
behind:The Legacy of Maximilian Kreuzer} edited by  Anton Rebhan, Ludmil Katzarkov,  Johanna Knapp, Radoslav Rashkov, Emanuel Scheid, World Scientific, 2013, arXiv:1012.2744.
\item{[25]} E. A. Ivanov and V. I. Ogievetsky, {\it The Inverse Higgs Phenomenon In Nonlinear Realizations}, Teor. Mat. Fiz.25 (1975) 164.
\item{[26]} E. Bergshoeff, E. Sezgin and P. Townsend, {\it Properties
of Êeleven dimensional supermembrane theory}, Ann. Phys. {\bf 185} (1988) 330;  E. Bergshoeff, E. Sezgin and P. Townsend, {\it Supermembranes and Eleven-Dimensional Supergravity}, Phys. Lett. {\bf 189} (1987) 75. 
\item{[27]}  A. Tumanov and P. West, {\it Generalised vielbeins and non-linear realisations },  arXiv:1405.7894. 
\item{[28]} P. West, {\it Introduction to Strings and Branes}, (2012), 
Cambridge University Press. 
\item{[29]} E. Bergshoeff and F. Riccioni, {\it D- Brane Wess-Zumino Terms and U-Duality}, ArXiv:10009.4657. 
\item{[30]} T.   Nutma,    SimpLie,    a   simple   program   for   Lie   algebras, https://code.google.com/p/simplie/.
\item{[31]} A. Kleinschmidt and P. West, {\it Representations of $G^{+++}$ and the role of  space-time}, JHEP, 0402 (2004) 033, hep-th/0312247. 
\item{[32]} S. Shiba and H. Sugawara, {\it M2- and M5-branes in E11 Current Algebra Formulation of M-theory },  arXiv:1709.07169. 
\item{[33]} M. Pettit and P. West.  to be published. 
\item{[34]} I. Schnakenburg and  P. West, {\it Kac-Moody   
symmetries of IIB supergravity}, Phys. Lett. {\bf B517} (2001) 421, hep-th/0107181.
\item{[35]} P. Howe, E. Sezgin and P. West,
{\it Covariant fieldÊÊequations of the M theory five-brane}, Phys. Lett.
{\bf 399B} (1997) 49, hep-th/9702008: I. Bandos, K Lechner, A. Nurmagambetov, P. Pasti and D.
Sorokin, and M. Tonin, {\it Covariant action for the super fivebrane of M-theory}, Phys.Rev.Lett. {\bf 78}  (1997) 4332, 
hep-th/9701149; M. Aganagic, J. Park, C. Popescu, and J. ÊSchwarz,
{\it Worldvolume action of the M-theory fivebrane},
hep-th/9701166.

\end